\documentclass[10pt,american,aps,prd,nofootinbib,superscriptaddress,preprintnumbers]{revtex4-2}
\usepackage[T1]{fontenc}
\usepackage[latin9]{inputenc}
\usepackage{amsmath}
\usepackage{amssymb}

\makeatletter
\def\Mpl{M_{\rm P}}
\usepackage{tikz}

\makeatother

\usepackage{babel}
\begin{document}
\preprint{YITP-22-59, IPMU22-0035}
\title{Extended minimal theories of massive gravity}
\author{Antonio De Felice}
\affiliation{Center for Gravitational Physics, Yukawa Institute for Theoretical
Physics, Kyoto University, 606-8502, Kyoto, Japan}
\author{Shinji Mukohyama}
\affiliation{Center for Gravitational Physics, Yukawa Institute for Theoretical
Physics, Kyoto University, 606-8502, Kyoto, Japan}
\affiliation{Kavli Institute for the Physics and Mathematics of the Universe (WPI),
The University of Tokyo, Kashiwa, Chiba 277-8583, Japan}
\author{Masroor C.\ Pookkillath}
\affiliation{Center for Gravitational Physics, Yukawa Institute for Theoretical
Physics, Kyoto University, 606-8502, Kyoto, Japan}
\date{\today}
\begin{abstract}
In this work, we introduce a class of extended Minimal Theories of Massive Gravity (eMTMG), without requiring a priori that the theory should admit the same homogeneous and isotropic cosmological solutions as the de Rham-Gabadadze-Tolley massive gravity. The theory is constructed as to have only two degrees of freedom in the gravity sector. In order to perform this step we first introduce a precursor theory endowed with a general graviton mass term, to which, at the level of the Hamiltonian, we add two extra constraints as to remove the unwanted degrees of freedom, which otherwise would typically lead to ghosts and/or instabilities. On analyzing the number of independent constraints and the properties of tensor mode perturbations, we see that the gravitational waves are the only propagating gravitational degrees of freedom which do acquire a non-trivial mass, as expected. In order to understand how the effective gravitational force works for this theory we then investigate cosmological scalar perturbations in the presence of a pressureless fluid. We then restrict the whole class of models by imposing the following conditions at all times: 1) it is possible to define an effective gravitational constant, $G_{{\rm eff}}$; 2) the value $G_{\text{eff}}/G_{N}$ is always finite but not always equal to unity (as to allow some non-trivial modifications of gravity, besides the massive tensorial modes); and 3) the square of mass of the graviton is always positive. These constraints automatically make also the ISW-effect contributions finite at all times. Finally we focus on a simple subclass of such theories, and show they already can give a rich and interesting phenomenology. 
\end{abstract}
\maketitle

\section{Introduction}

In these last years, we have witnessed a boom for the research in
gravity both from theoretical and experimental sides. In particular,
the discovery of gravitational waves has paved ground for a long research
path which will lead to a deeper understanding of several new aspects
of gravity~\citep{LIGOScientific:2016aoc}. On one side this will
affect largely astrophysics and in particular the research aimed to
understand the dynamics of the final states of stars in strong gravity
regimes. On another end, a large sample from the detected gravitational
waves seems to be coming from the mergers of two black holes: the
values for the masses of the black holes involved in these phenomena
seem to be pointing either to non-trivial astrophysical sources or
even to the existence of primordial black holes, which could be forming
at least part of the dark matter content~\citep{Carr:2016drx}.

From an observational point of view, having a larger sample of neutron
stars mergers will also give us a link to cosmology, since the sources
of the signals could be located in a far away galaxy, leading to a
propagation of the gravitational waves over a cosmological distance~\citep{Schutz:1986gp}.
In particular, this branch of the gravitational wave science should help
us understanding the nature of the so-called $H_{0}$ tension~\citep{Bernal:2016gxb,DiValentino:2021izs}.
As a matter of fact, the high redshift CMB data including Planck~\citep{Planck:2018vyg}
as well as Atacama Cosmology Telescope (ACT)~\citep{ACT:2020frw},
and the late time data, SH0ES~\citep{Riess:2021jrx} do not agree
with each other in the context of $\Lambda$CDM, the ``de facto''
standard model of gravity. This tension could point either to new
physics or to some unexpected and non-trivial systematic errors in
the data, and the gravitational waves discoveries should help confirming
or ruling out this last hypothesis.

If this situation is not already surprising, in cosmology, still
another observable in the data, related to the growth of structure,
the amplitude of the fluctuation $S_{8}$, during matter domination
up to now, seems to be again showing not good agreement between early-time
data (Planck~\citep{Planck:2018vyg}) and late-time Large Scale Structures~\citep{DiValentino:2020vvd,DES:2021bvc},
once more, in the context of the $\text{\ensuremath{\Lambda}CDM model}$.
These two tensions open up a room for exploring models of universe beyond
the $\Lambda$CDM, for example by modifying gravity at large scales~\citep{Clifton:2011jh,DeFelice:2020prd,DeFelice:2020sdq}.
See~\citep{DiValentino:2021izs} for a review of possible solutions
to the Hubble tension.

There have been several attempts to try to reconcile data and theory
at the cost of introducing new degrees of freedom, which could change the dynamics of the cosmological background and matter perturbation needed
to solve the above mentioned puzzling tensions~\citep{DeFelice:2020sdq,Amendola:2020ldb,DeFelice:2020icf,VCDM:solvingH0,DeFelice:2020prd}.
What is surprising though is that at local scales (e.g.\ solar system
scales) there is no trace of such additional degrees of freedom which
would be necessary to fix the cosmological issues~\citep{Will:2014kxa}.
One then needs to address how to hide existing new degrees of freedom
in environments with energy scales much higher than the cosmological ones~\citep{Joyce:2014kja,Brax:2015cla,Koyama:2015vza}.

However, a more minimal approach, and possibly simpler one, is to
give a non-zero mass to the graviton~\citep{Fierz_Pauli}. If
the mass, $\mu$, of such a mode is small enough, i.e.\ comparable
to the size of today's Hubble expansion rate ($\mu\simeq10^{-33}$ eV), then
for the typical energy scales present in astrophysical environments,
the graviton would typically be largely ultra-relativistic avoiding
in this way the constraints on $\mu$ coming from the propagation
of gravitational waves, which is $\mu<10^{-23}$ eV~\citep{LIGOScientific:2017bnn}.
Even though the graviton mass is negligible at very short scales (i.e.\ solar
system scales), at cosmological scales things could be different.
In fact, the theory leading to a non-zero graviton mass could be becoming
sensibly different from $\Lambda$CDM at late times, when $H\simeq\mu$,
and this theory could be responsible for an apparent modified gravity
behavior in cosmology which could be affecting both the background
and cosmological perturbations, being able in this way to address
both the above mentioned tensions~\citep{deAraujo:2021cnd,DeFelice:2021trp}.

Is this an interesting idea or nothing but a theorist-wild-dream scenario?
In fact, the question of a non-zero mass for the graviton has been
posed long time ago and first partially addressed by Fierz and Pauli~\citep{Fierz_Pauli}.
Partially, because they studied a theory of massive gravity only in
a perturbative regime, i.e.\ without knowing the theory in full,
in any non-perturbative regime. Only quite recently, a theory of massive
gravity which is totally consistent from a theoretical point of view,
was introduced, which is dubbed as dRGT theory~\citep{dRGT_1,dRGT_2}.
This breakthrough led to an exploration of the phenomenology for such
a theory, but it was realized that this model, at least in the simplest
approach, could not be having a well-defined cosmological behavior~\citep{dRGT_no_FLRW,DeFelice:2013awa}.
By a beyond-linear-perturbation analysis around a homogeneous and isotropic
background, it was found that at least one (out of five) of the graviton degrees of freedom
would be a (light) ghost and as such would make dRGT loose its ability
to make predictions \citep{dRGT_no_FLRW}.

Although this result might look disappointing, this negative result
has led to several other possibilities. One of them consisted of introducing
terms which break Lorentz invariance, in order to remove unwanted
(unstable) degrees of freedom. Along these lines of research, a model
called minimal theory of massive gravity (MTMG) was introduced as
to resolve the issue of dRGT on a cosmological background~\citep{MTMG:origpap}.
In particular MTMG, by construction, removes three (out of five) graviton degrees of freedom in a non-linear
way, leaving tensor modes as only propagating degrees of freedom on any background. The theory has been proved to be
interesting and was leading to a non-trivial phenomenology discussed
even recently in the literature~\citep{MTMG:pheno,deAraujo:2021cnd,DeFelice:2021trp}.
Along the same lines, MTMG was extended as to have a scalar field
in the gravity sector (in addition to the massive graviton)~\citep{MQD_OP,MQD_Horndeski,MQD_pheno},
and even to a minimal theory of bigravity (MTBG)~\citep{DeFelice:2020ecp}.

Besides the requirement for the minimal number of propagating degrees of freedom, MTMG has been constructed so as to admit the same homogeneous and isotropic cosmological solutions as in dRGT, for which there are two branches of solutions: the self-accelerating branch and the normal branch. In the self-accelerating branch the graviton mass term acts as an effective cosmological constant that can accelerate the expansion of the universe~\citep{Gumrukcuoglu:2011ew} while the linear perturbations behave exactly the same as the standard $\Lambda$CDM except that gravitational waves acquires a non-vanishing mass. Unlike dRGT, the self-accelerating branch of MTMG is free from fatal instabilities and thus provides a firm testing ground for gravitational wave physics of massive gravity. However, from the viewpoint of recent tensions in cosmology, this branch of MTMG is as good as but not better than $\Lambda$CDM. In this respect the normal branch of MTMG could perform better than $\Lambda$CDM. Indeed, in the normal branch of MTMG the scalar linear perturbations behave differently from $\Lambda$CDM.

Although the normal branch of MTMG has proved to be an interesting possibility as to try to modify gravity at large scales in a consistent and minimal way, still it had some features which were setting some theoretical and phenomenological issues. In particular, MTMG was leading to a modified effective Newtonian gravitational constant which at large scales behaves as $G_{{\rm eff}}/G_{N}\propto(\mu^{2}/H^{2}-2)^{-2}$~\citep{MTMG:pheno}. This expression for $G_{{\rm eff}}/G_{N}$ is well-behaved for negative-squared-mass for the graviton (for which, though, a tachyonic instability, with a time-scale of order $H_{0}^{-1}$, would be affecting modes of order $k/(a_{0}H_{0})\simeq1$)~\citep{MTMG:isw}. But for a large-enough (but still inside the allowed Ligo bounds) positive-squared-mass graviton, a range of positive $\mu^{2}$, for $\mu\simeq2H_{0},$ would lead to strong modifications to $G_{{\rm eff}}/G_{N}$, leading in turn to strong constraints from the data even at non-linear scales \citep{Hagala:2020eax}. In particular, in a recent paper, on studying the effect of Planck data on MTMG, it was discovered that positive $\mu^{2}$ is actually preferred but because of the above mentioned behavior of $G_{{\rm eff}}/G_{N}$, $\mu^{2}$ is strongly constrained toward values very close to zero~\citep{DeFelice:2021trp}. This phenomenon puts strong limits on the normal branch of MTMG.

In this paper, we try to solve these issues of the normal branch of MTMG by extending the MTMG itself, in a way which is meant to cure the above mentioned behavior of $G_{{\rm eff}}/G_{N}$. In order to extend MTMG we still need to add constraints to the Hamiltonian of a precursor theory as to remove the unwanted degrees of freedom, but we change the constraints themselves. One of the constraint of MTMG was chosen as to admit exactly the same cosmological background as dRGT. As mentioned above, this constraint defining MTMG was leading to the presence of two branches for the background dynamics. On the other hand, the extended MTMG (eMTMG) has in general a different background dynamics from dRGT, especially if these same modifications/extensions lead to a better behaved phenomenology. Indeed, eMTMG allows for a much larger freedom in terms of background dynamics, still being 
a minimal theory (i.e.\ with only two tensor propagating degrees of freedom on any background). However, as we shall see later on, the condition that on any allowed cosmological background $G_{{\rm eff}}/G_{N}$ will never have poles and $\mu^{2}$ being non-negative, will considerably reduce the set of allowed theories. Still, we give a proof of existence of a large class of models which indeed satisfy these criteria (and which by construction does not reduce to dRGT at the background level). We also show that at the level of the background and linear perturbations, all predictions of the models in this class are captured by a smaller subclass of eMTMG with only 6 parameters, which determine the cosmological constant and the behavior of $\mu_{0}$ and $G_{{\rm eff,0}}/G_{N},$. As was happening in MTMG, we find that for environmental densities much larger than the present cosmological ones, that is $\rho\gg\Mpl^{2}H_{0}^{2}$ (valid at solar system scales and at high redshifts) we find that $G_{{\rm eff}}/G_{N}\to1$.

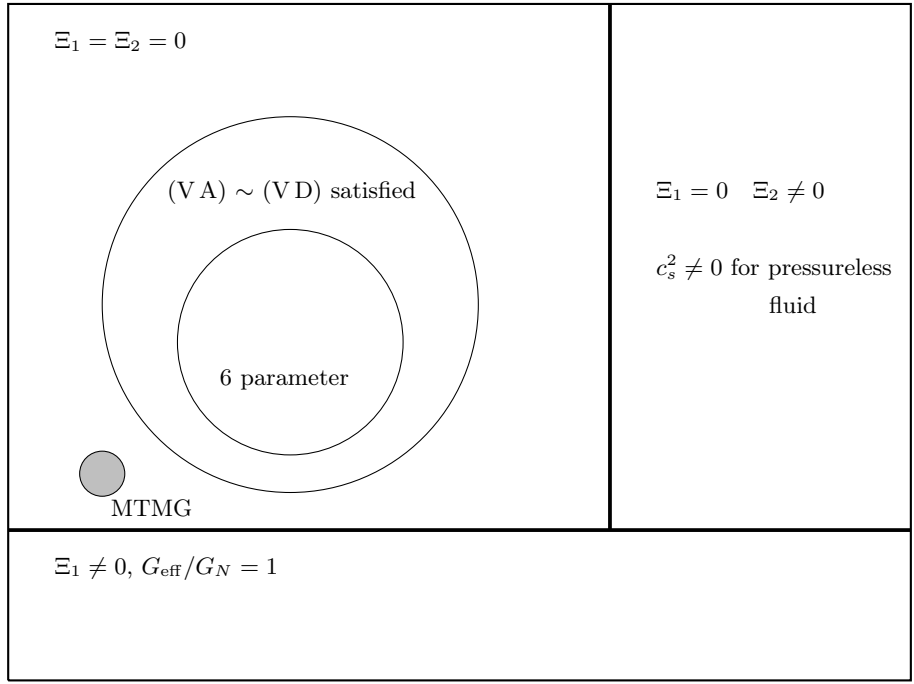
\begin{figure}[ht]
\begin{tikzpicture}
\draw[thick] (0,0) -- (12,0) -- (12,9) -- (0,9) -- (0,0);
\draw (3.75,5) circle (2.5cm);
\draw (3.75,4.5) circle (1.5cm);
\draw[fill=lightgray] (1.25,2.75) circle (0.3cm);
\draw[line width=0.5mm, black] (8,2) -- (8,9);
\draw[line width=0.5mm, black] (0,2) -- (12,2);
\filldraw[black] (0.5,8.5)   node[anchor=west]{$\Xi_{1}=\Xi_{2}=0$};
\filldraw[black] (8.5,6.5)   node[anchor=west]{$\Xi_{1} = 0 \quad \Xi_{2} \neq 0$};
\filldraw[black] (8.5,5.5)   node[anchor=west]{$c_{s}^{2}\neq0$ for pressureless};
\filldraw[black] (10.,5.)   node[anchor=west]{fluid};
\filldraw[black] (0.5,1.5)   node[anchor=west]{$\Xi_{1} \neq 0$, $G_{\rm eff}/G_N=1$};
\filldraw[black] (1.25,2.3)   node[anchor=west]{MTMG};
\filldraw[black] (2.0,6.5)   node[anchor=west]{(\ref{subsec:criteronGefftoGN@earlytime}) $\sim$ (\ref{subssec:ISW_finite}) satisfied};
\filldraw[black] (2.7,4)   node[anchor=west]{6 parameter};
\end{tikzpicture} 

\caption{This figure shows relations among different subclasses of the extended
MTMG (eMTMG) and the original MTMG. The classification has been made according
to two scalar quantities which determine the phenomenology of $G_{{\rm eff}}$.
Other criteria for the classification can be in principle considered.
The region where phenomenological criteria (\ref{subsec:criteronGefftoGN@earlytime}) $\sim$ (\ref{subssec:ISW_finite}) are satisfied determines
a class of models with appealing phenomenological properties, e.g.\ $G_{{\rm eff}}/G_{N}$
is finite for any dynamics of the cosmological background, the tensor graviton
has a non-negative mass squared, etc. Finally it is possible to give a simple
subset (having at most six free parameters) which already possesses
all the defining properties of the model.}
\end{figure}

This paper is organized as follows. Section~\ref{sec:model} shows
the construction of the eMTMG, where we introduce two general
functions, $F_{1}$ and $F_{2}$, for which we make use of the Cayley-Hamilton
theorem. In particular, after writing down a precursor theory, we
add constraints to make the theory minimal, in the sense that no additional
degrees of freedom are propagating in the gravitational sector besides
the gravitational waves, which become massive. Then, in section~\ref{sec:background},
we study the spatially flat, homogeneous and isotropic cosmological
background in this theory. Here, using the minisuperspace Hamiltonian
and the constraints, we show that one of the Lagrange multipliers
$\lambda(t)$ vanishes in the spatially flat, homogeneous ans isotropic
background. Unlike the original MTMG, the space of solutions is not
separated into two branches: the self-accelerating branch and the
normal branch, rather there is one and only one universal branch.
While in Appendix~\ref{app:selfacce} we consider the condition under
which the separation into the two branches occurs, in the rest of
the present paper we study the general case. In section~\ref{sec:linearperturbations},
we study linear perturbations around the spatially flat, homogeneous
and istotropic background in this theory. We first consider the propagation
of the gravitational waves on the cosmological background. As expected,
the two modes are now massive. Subsequently, we derive the expression
for the $G_{\text{eff}}/G_{N}$ considering the eMTMG minimally
coupled with a pressure-less fluid. Furthermore, we derive equations
of motion for scalar perturbations in the presence of multiple perfect
fluids with general equations of state. In section~\ref{sec:criteria}
we then make a list of phenomenologically motivated criteria to be
imposed on the theory, which makes it possible for us to find a subset
of models with a finite number of parameters. In particular, we require
the finiteness at any redshift of $G_{\text{eff}}/G_{N}$ which is
anyhow modified at late times, i.e.\ without altering the early time
dynamics. In addition, we impose the condition that the squared
mass of the gravitational waves is positive, i.e.\ $\mu^{2}>0$.
We also demand the finiteness of the ISW effect at any redshift. In
order to give a working example for such a theory, which is nonetheless
endowed with the desired features of the general model, in section~\ref{sec:polynomialansatz}
we adopt a simple polynomial ansatz for $F_{1,2}$ and impose the
phenomenological criteria explained above step by step. As a consequence,
we obtain a rather simple subclass of the general model which satisfies
all the criteria. It turns out that at the level of the background
and the linear perturbations, all observables within this subclass
depend only on six parameters while $F_{1,2}$ depend on more parameters.
We thus remove this degeneracy by defining a further simpler subclass,
by picking up the model for $F_{1,2}$ which only shows the above-mentioned
six free parameters, i.e.\ five more than $\Lambda$CDM. Finally,
we report our conclusion in section~\ref{sec:concl}. We find it
useful to add five appendices to the main text. Appendix~\ref{app:variations}
shows some useful variational formulae needed for the construction
of the theory. In Appendix~\ref{app:MTMG-subcase} we discuss the
original MTMG as a special case of this eMTMG. In Appendix~\ref{app:selfacce}
we consider the condition under which the space of spatially flat,
homogeneous and isotropic solutions of the eMTMG is divided
into two branches, the so-called self-accelerating and normal branches.
In Appendix~\ref{app:Geff_GN}, we provide the full expression for
$G_{\text{eff}}/G_{N}$ and the ISW potential field. Finally, Appendix~\ref{app:masslesscase}
discusses a rather peculiar model having massless tensor modes (on
the cosmological background), with a non-trivial dynamics for the
scalar perturbations, i.e.\ $G_{{\rm eff}}/G_{N}\neq1$.

\section{Model construction}

\label{sec:model}

\subsection{Building blocks}

In order to build up the model, we will follow a path which is similar
to the one followed in \citep{MTMG:origpap,MTMG:pheno}. First of
all, in the following, we will make use of the unitary gauge and the
metric formalism\footnote{Using the unitary gauge, although not strictly necessary turns out
to be simplifying the calculations. As for the choice of the metric
formalism, one could equivalently choose the vielbein formalism to
define the theory, as done in \citep{MTMG:origpap,MTMG:pheno}.}. In the unitary gauge we introduce a three-dimensional fiducial metric
with positive definite signature, which is, by construction of the
theory, an external, explicitly time (and time only) dependent field,
that we denote by $\tilde{\gamma}_{ij}(t)$. In the unitary gauge
we will also introduce another external field, $M$, that we call
fiducial lapse function. In order for the theory to allow spatially
flat, homogeneous and isotropic solutions, we require the fiducial
sector to be compatible with the symmetry of such solutions. For this
reason we will identify $\tilde{\gamma}_{ij}=\tilde{a}(t)^{2}\,\delta_{ij}$
and $M=M(t)$, where $\tilde{a}(t)$ is the fiducial scale factor.
This three-dimensional fiducial metric admits an inverse, denoted
by $\tilde{\gamma}^{ij}$, which satisfies $\tilde{\gamma}^{il}\tilde{\gamma}_{lj}=\delta^{i}{}_{j}$.
Out of these external fields, we can also define the following field
$\bar{\zeta}^{i}{}_{j}$, as 
\begin{equation}
\bar{\zeta}^{i}{}_{j}\equiv\frac{1}{2M}\,\tilde{\gamma}^{il}\dot{\tilde{\gamma}}_{lj}\,,
\end{equation}
which describes the rate of change of the fiducial metric\footnote{In the vielbein formalism we instead define $\tilde{\zeta}^{i}{}_{j}=\frac{1}{M}\,E^{i}{}_{A}\,\dot{E}^{A}{}_{j}$,
where $\tilde{\gamma}_{ij}=\delta_{AB}\,E^{A}{}_{i}\,E^{B}{}_{j}$,
giving $\bar{\zeta}^{i}{}_{j}=\frac{1}{2}\,(\tilde{\zeta}^{i}{}_{j}+\tilde{\gamma}^{il}\,\tilde{\zeta}^{k}{}_{l}\,\tilde{\gamma}_{jk})$,
which in any case agree with each other when $\tilde{\gamma}_{ij}=\tilde{a}^{2}\,\delta_{ij}$.}. Notice that in the unitary gauge description, having the presence of the external fields which required a full coordinate choice, will explicitly break four-dimensional diffeomorphism, and a choice of slicing has been automatically fixed.

Of course, we also have physical, dynamical metric variables, which
we adopt from the ADM formalism. In particular we have a lapse function
$N$, a shift vector $N^{i}$ and a three-dimensional metric $\gamma_{ij}$,
which admits the inverse $\gamma^{ij}$. Out of them, the four-dimensional
physical metric can be written as 
\begin{equation}
g_{\mu\nu}dx^{\mu}dx^{\nu}=-N^{2}dt^{2}+\gamma_{ij}(dx^{i}+N^{i}dt)(dx^{j}+N^{j}dt).\label{eqn:4dmetric-ADM}
\end{equation}
Having introduced the dynamical field $\gamma_{ij}$ and the external
field $\tilde{\gamma}_{ij}$, which in unitary gauge has a fixed,
given dynamics, we can introduce the building blocks of the theory
$\mathcal{K}^{i}{}_{j}$ and $\mathfrak{K}^{i}{}_{j}$ which satisfy
the following properties 
\begin{eqnarray}
\mathcal{K}^{i}{}_{l}\mathcal{K}^{l}{}_{j} & = & \tilde{\gamma}^{il}\gamma_{lj}\,,\\
\mathfrak{K}^{i}{}_{l}\mathfrak{K}^{l}{}_{j} & = & \gamma^{il}\tilde{\gamma}_{lj}\,,\\
\mathcal{K}^{i}{}_{l}\mathfrak{K}^{l}{}_{j} & = & \delta^{i}{}_{j}=\mathfrak{K}^{i}{}_{l}\mathcal{K}^{l}{}_{j}\,.
\end{eqnarray}
Some useful formulae for the variations of the quantities defined
above are summarized in Appendix~\ref{app:variations}.

\subsection{Precursor Hamiltonian}

We now have all the required building blocks to define the theory,
and we will do so by writing down its Hamiltonian density, and then
via a Legendre transformation, we will find its Lagrangian density.
Then, along the same lines of MTMG, see e.g.\ \citep{MTMG:pheno},
we first introduce a precursor Hamiltonian density, which we now define
to be 
\begin{equation}
\mathcal{H}_{{\rm pre}}\equiv-N\,\mathcal{R}_{0}^{{\rm GR}}-N^{i}\mathcal{R}_{i}+\frac{1}{2}\,m^{2}\Mpl^{2}N\sqrt{\gamma}\,F_{1}([\mathfrak{K}],[\mathfrak{K}^{2}],[\mathfrak{K}^{3}])+\frac{1}{2}\,m^{2}\Mpl^{2}M\sqrt{\tilde{\gamma}}\,F_{2}([\mathcal{K}],[\mathcal{K}^{2}],[\mathcal{K}^{3}])\,,\label{eq:H_pre}
\end{equation}
where 
\begin{eqnarray}
\mathcal{R}_{0}^{{\rm GR}} & = & \frac{\Mpl^{2}}{2}\sqrt{\gamma}\,R^{(3)}-\frac{2}{\Mpl^{2}}\sqrt{\gamma}\left(\gamma_{ik}\gamma_{jl}-\frac{1}{2}\gamma_{ji}\gamma_{kl}\right)\tilde{\pi}^{ij}\tilde{\pi}^{lk}\,,\\
\mathcal{R}_{i} & = & 2\sqrt{\gamma}\gamma_{ij}D_{k}\tilde{\pi}^{jk}\,,\\
\tilde{\pi}^{ij} & \equiv & \frac{\pi^{ij}}{\sqrt{\gamma}}\,,
\end{eqnarray}
and $[\mathcal{K}]\equiv\mathcal{K}^{i}{}_{i},$ $[\mathcal{K}^{2}]\equiv\mathcal{K}^{i}{}_{j}\,\mathcal{K}^{j}{}_{i}$,
etc. Here we point out that the fields $N$ and $N^{i}$ have been
considered to be Lagrange multipliers, whereas the dynamical degrees
of freedom enter in the six independent components of $\gamma_{ij}$,
which lead, in turn, to twelve phase-space variables, since $\pi^{ij}$
correspond to their conjugate momenta. Here the operator $D_{i}$
represents the covariant derivative compatible with the three dimensional
metric $\gamma_{ij}$.

By looking at Eq.\ (\ref{eq:H_pre}), the precursor theory is defined
in terms of two functions $F_{1,2}$, which depend on the trace of
powers of the above defined building blocks $\mathcal{K}^{i}{}_{j}$
and $\mathfrak{K}^{i}{}_{j}$. Making use of the Cayley-Hamilton theorem
applied to a three-dimensional matrix, e.g.\ $\mathfrak{K}^{i}{}_{j}$,
we only choose $[\mathfrak{K}],[\mathfrak{K}^{2}],[\mathfrak{K}^{3}]$
as the variables out of which the function $F_{1}$ depends on. Also
by the same theorem, the mirror variables $[\mathcal{K}],[\mathcal{K}^{2}],[\mathcal{K}^{3}]$
can be rewritten in terms of the previous $[\mathfrak{K}],[\mathfrak{K}^{2}],[\mathfrak{K}^{3}]$
variables, which become the really independent ones. Therefore, on
looking at the precursor Hamiltonian, we can further define the following
two quantities 
\begin{eqnarray}
\mathcal{R}_{0} & \equiv & \mathcal{R}_{0}^{{\rm GR}}-\frac{1}{2}\,m^{2}\Mpl^{2}\sqrt{\gamma}\,F_{1}([\mathfrak{K}],[\mathfrak{K}^{2}],[\mathfrak{K}^{3}])\,,\\
H_{1} & = & \frac{1}{2}\,m^{2}\Mpl^{2}\int d^{3}xM\sqrt{\tilde{\gamma}}\,F_{2}([\mathcal{K}],[\mathcal{K}^{2}],[\mathcal{K}^{3}])\,.
\end{eqnarray}
Indeed, for this precursor theory, the four Lagrange multipliers $N$
and $N^{i}$ set four constraints, whereas $H_{1}$ corresponds to
the Hamiltonian of the precursor theory evaluated on the constraint
surface (on which $\mathcal{R}_{0}$ and $\mathcal{R}_{i}$ all vanish).
One can then evaluate the time derivative of the constraints $\mathcal{R}_{0}$
and $\mathcal{R}_{i}$. As for $\dot{\mathcal{R}}_{0}$, we would find 
$\dot{\mathcal{R}}_{0}=-N^{i}\{\mathcal{R}_{0},\mathcal{R}_{i}\}+\dots$,
which needs to vanish on the constraints surface. However the 
Poisson brackets $\{\mathcal{R}_{0},\mathcal{R}_{i}\}$ do not
all vanish, then setting $\dot{\mathcal{R}}_{0}\approx0$, would actually
fix one of the Lagrange multipliers without imposing any new constraint
on the theory. Indeed, since the rank of $\{\mathcal{R}_{0},\mathcal{R}_{i}\}$
is two, not all the eight $\mathcal{R}_{0}$, $\mathcal{R}_{i}$,
$\dot{\mathcal{R}}_{0}$, $\dot{\mathcal{R}}_{i}$ are constraints,
but only six of them. This means that this theory has $\tfrac{1}{2}\,(12-6)=3$
degrees of freedom, where twelve represents the number of independent
components of $\gamma_{ij}$ and their conjugate momenta in the phase
space.

\subsection{Hamiltonian of the extended minimal theory of massive gravity}

From what we have learned in the previous section, we still need
to add two new constraints as to make the theory minimal, i.e.\ having
only two propagating degrees of freedom in the gravity sector. In
order to achieve this goal, we can follow the same steps of MTMG as
to make the theory minimal. Let us use the on-shell precursor Hamiltonian
$H_{1}$ as to define the quantities $\mathcal{C}_{0}$ and $\mathcal{C}_{i}$ as follows. 
They would correspond to time-derivatives of the $\mathcal{R}_{0}$ and 
$\mathcal{R}_{i}$ constraints if $H_1$ were the Hamiltonian of the system. 
\begin{eqnarray}
\mathcal{C}_{0} & \equiv & \{\mathcal{R}_{0},H_{1}\}+\frac{\partial\mathcal{R}_{0}}{\partial t}\nonumber \\
 & = & m^{2}M\sqrt{\tilde{\gamma}}(2\tilde{\gamma}^{ij}\tilde{\pi}^{ce}-\tilde{\gamma}^{ic}\tilde{\pi}^{je})\left[\frac{1}{2}F_{2,[\mathcal{K}]}\,\mathfrak{K}^{l}{}_{i}\gamma_{lc}\gamma_{je}+F_{2,[\mathcal{K}^{2}]}\,\gamma_{ic}\gamma_{je}+\frac{3}{2}F_{2,[\mathcal{K}^{3}]}\,\mathcal{K}^{l}{}_{i}\gamma_{lc}\gamma_{je}\right]\nonumber \\
 & - & m^{2}\Mpl^{2}\sqrt{\gamma}\,M\bar{\zeta}^{i}{}_{j}\left[\frac{1}{2}F_{1,[\mathfrak{K}]}\,\mathcal{K}^{l}{}_{k}\tilde{\gamma}_{li}\gamma^{kj}+F_{1,[\mathfrak{K}^{2}]}\,\tilde{\gamma}_{il}\gamma^{jl}+\frac{3}{2}F_{1,[\mathfrak{K}^{3}]}\,\mathfrak{K}^{j}{}_{l}\,\tilde{\gamma}_{ik}\gamma^{lk}\right],\\
\mathcal{C}_{i}[v^{i}] & = & \int d^{3}x\mathcal{C}_{i}v^{i}\equiv\{\mathcal{R}_{i}[v^{i}],H_{1}\}=\left\{ \int d^{3}x\mathcal{R}_{i}v^{i},H_{1}\right\} \nonumber \\
 & = & \frac{1}{2}m^{2}\Mpl^{2}\int d^{3}x\sqrt{\gamma}D_{j}v^{i}\left[\frac{1}{2}M\,\frac{\sqrt{\tilde{\gamma}}}{\sqrt{\gamma}}\,F_{2,[\mathcal{K}]}\,(\mathfrak{K}^{j}{}_{l}\tilde{\gamma}^{lk}\gamma_{ik}+\mathfrak{K}^{k}{}_{l}\tilde{\gamma}^{lj}\,\gamma_{ki})\right.\nonumber \\
 & + & \left.2M\,\frac{\sqrt{\tilde{\gamma}}}{\sqrt{\gamma}}\,F_{2,[\mathcal{K}^{2}]}\,\gamma_{il}\tilde{\gamma}^{jl}\frac{3}{2}M\,\frac{\sqrt{\tilde{\gamma}}}{\sqrt{\gamma}}\,F_{2,[\mathcal{K}^{3}]}\,(\mathcal{K}^{j}{}_{l}\tilde{\gamma}^{lk}\gamma_{ik}+\mathcal{K}^{k}{}_{l}\tilde{\gamma}^{lj}\gamma_{ki})\right],
\end{eqnarray}
where we have taken into consideration the fact that having chosen the unitary
gauge, the constraint $\mathcal{R}_{0}$ explicitly depends on time.
The previous relations lead to 
\begin{eqnarray}
\mathcal{C}_{0} & = & \frac{1}{2}m^{2}\Mpl^{2}M\sqrt{\tilde{\gamma}}\,(2\tilde{\gamma}^{bd}\tilde{\pi}^{ce}-\tilde{\gamma}^{bc}\tilde{\pi}^{de})[F_{2,[\mathcal{K}]}\,\mathfrak{K}^{a}{}_{b}\gamma_{ac}\gamma_{de}+2F_{2,[\mathcal{K}^{2}]}\,\gamma_{bc}\gamma_{de}+3F_{2,[\mathcal{K}^{3}]}\,\mathcal{K}^{a}{}_{b}\gamma_{ac}\gamma_{de}]+\sqrt{\gamma}\mathcal{C_{\zeta}}\,,\\
\mathcal{C}_{\zeta} & \equiv & -\frac{1}{2}m^{2}\Mpl^{2}M\,\bar{\zeta}^{c}{}_{d}\,(F_{1,[\mathfrak{K}]}\,\gamma^{db}\mathcal{K}^{a}{}_{b}\tilde{\gamma}_{ac}+2F_{1,[\mathfrak{K}^{2}]}\,\gamma^{db}\tilde{\gamma}_{bc}+3F_{1,[\mathfrak{K}^{3}]}\,\mathfrak{K}^{d}{}_{b}\,\gamma^{be}\tilde{\gamma}_{ec})\,,\\
\mathcal{C}_{i} & = & -m^{2}\Mpl^{2}\sqrt{\gamma}D_{j}\!\left\{ M\frac{\sqrt{\tilde{\gamma}}}{\sqrt{\gamma}}\!\left[\frac{F_{2,[\mathcal{K}]}}{4}(\mathfrak{K}^{j}{}_{l}\tilde{\gamma}^{lk}+\mathfrak{K}^{k}{}_{l}\tilde{\gamma}^{lj})\gamma_{ki}+F_{2,[\mathcal{K}^{2}]}\tilde{\gamma}^{jl}\gamma_{il}+\frac{3}{4}F_{2,[\mathcal{K}^{3}]}(\mathcal{K}^{j}{}_{l}\tilde{\gamma}^{lk}+\mathcal{K}^{k}{}_{l}\tilde{\gamma}^{lj})\gamma_{ki}\right]\right\} \!.
\end{eqnarray}
We are now ready to define the extended-MTMG theory by giving its
Hamiltonian density as 
\begin{equation}
\mathcal{H}=-N\,\mathcal{R}_{0}-N^{i}\mathcal{R}_{i}+\frac{1}{2}m^{2}\Mpl^{2}M\sqrt{\tilde{\gamma}}\,F_{2}([\mathcal{K}],[\mathcal{K}^{2}],[\mathcal{K}^{3}])-\lambda\mathcal{C}_{0}-\lambda^{i}\mathcal{C}_{i}\,.
\end{equation}
Now all the eight constraints, imposed by the Lagrange multipliers
$N$, $N^{i}$, $\lambda$, $\lambda^{i}$, are second class which
then leave only two dynamical degrees of freedom. We can write down
the Hamiltonian of the theory as 
\begin{equation}
H=\int d^{3}x\,[-N\,\mathcal{R}_{0}-N^{a}\mathcal{R}_{a}+\frac{1}{2}m^{2}\Mpl^{2}M\sqrt{\tilde{\gamma}}\,F_{2}([\mathcal{K}],[\mathcal{K}^{2}],[\mathcal{K}^{3}])-\lambda\mathcal{C}_{0}-\sqrt{\gamma}\,(D_{j}\lambda^{i})\mathcal{C}^{j}{}_{i}],
\end{equation}
where we have introduced the three-dimensional tensor 
\begin{equation}
\mathcal{C}^{j}{}_{i}\equiv\frac{1}{2}m^{2}\Mpl^{2}\,M\,\frac{\sqrt{\tilde{\gamma}}}{\sqrt{\gamma}}\left[\frac{1}{2}\,F_{2,[\mathcal{K}]}\,(\mathfrak{K}^{j}{}_{k}\tilde{\gamma}^{kl}+\tilde{\gamma}^{jk}\,\mathfrak{K}^{l}{}_{k})\gamma_{li}+2F_{2,[\mathcal{K}^{2}]}\,\tilde{\gamma}^{jk}\gamma_{ki}+\frac{3}{2}\,F_{2,[\mathcal{K}^{3}]}\,(\mathcal{K}^{j}{}_{k}\tilde{\gamma}^{kl}+\tilde{\gamma}^{jk}\mathcal{K}^{l}{}_{k})\gamma_{li}\right].
\end{equation}
In summary, since the constraints for the theory now add to eight,
the theory is minimal, i.e.\ the number of gravitational degrees
of freedom is now $\frac{1}{2}\,(12-8)=2$.

\subsection{Minimal theory Lagrangian}

In order to find the Lagrangian density of the theory, we need to
perform a Legendre transformation. From the Hamiltonian equations
of motion for $\gamma_{ij}$, we find 
\begin{eqnarray}
\dot{\gamma}_{ij} & = & \{\gamma_{ij},H_{{\rm tot}}\}=\frac{2N}{\Mpl^{2}}\,(2\gamma_{ik}\gamma_{jd}-\gamma_{ij}\gamma_{kd})\,\tilde{\pi}^{kd}+\gamma_{ik}D_{j}N^{k}+\gamma_{jk}D_{i}N^{k}\nonumber \\
 & + & \frac{1}{2}m^{2}\lambda M\,\frac{\sqrt{\tilde{\gamma}}}{\sqrt{\gamma}}\,[2\tilde{\gamma}^{kd}F_{2,[\mathcal{K}^{2}]}(\gamma_{ij}\gamma_{kd}-2\gamma_{ik}\gamma_{jd})-(\mathfrak{K}^{k}{}_{d}F_{2,[\mathcal{K}]}+3\mathcal{K}^{k}{}_{d}F_{2,[\mathcal{K}^{3}]})\tilde{\gamma}^{de}(\gamma_{ie}\gamma_{jk}+\gamma_{ik}\gamma_{je}-\gamma_{ij}\gamma_{ke})]\,,
\end{eqnarray}
so that we can also find the relation between the extrinsic curvature
$K_{ij}\equiv\frac{1}{2N}\,(\dot{\gamma}_{ij}-\gamma_{ik}D_{j}N^{k}-\gamma_{jk}D_{i}N^{k})$
and the canonical momenta $\pi^{ij}$ as 
\begin{eqnarray}
K_{ij} & = & \frac{1}{\Mpl^{2}}\,(2\gamma_{ik}\gamma_{jd}-\gamma_{ij}\gamma_{kd})\,\tilde{\pi}^{kd}\nonumber \\
 & + & \frac{m^{2}}{4}\lambda\frac{M}{N}\,\frac{\sqrt{\tilde{\gamma}}}{\sqrt{\gamma}}\,[2\tilde{\gamma}^{kd}F_{2,[\mathcal{K}^{2}]}(\gamma_{ij}\gamma_{kd}-2\gamma_{ik}\gamma_{jd})-(\mathfrak{K}^{k}{}_{d}F_{2,[\mathcal{K}]}+3\mathcal{K}^{k}{}_{d}F_{2,[\mathcal{K}^{3}]})\tilde{\gamma}^{de}(\gamma_{ie}\gamma_{jk}+\gamma_{ik}\gamma_{je}-\gamma_{ij}\gamma_{ke})]\,,
\end{eqnarray}
out of which we have 
\begin{equation}
\tilde{\pi}^{ij}=\frac{\Mpl^{2}}{2}\,(\gamma^{ik}\gamma^{jd}-\gamma^{ij}\gamma^{kd})K_{kd}-\frac{m^{2}\Mpl^{2}}{8}\,\frac{M}{N}\,\lambda\,\Theta^{ij}\,.
\end{equation}
Here, we have introduced the following tensor 
\begin{equation}
\Theta^{ij}=-\frac{\sqrt{\tilde{\gamma}}}{\sqrt{\gamma}}\left[(\tilde{\gamma}^{jk}\mathfrak{K}^{i}{}_{k}+\tilde{\gamma}^{ik}\mathfrak{K}^{j}{}_{k})F_{2,[\mathcal{K}]}+4\tilde{\gamma}^{ij}F_{2,[\mathcal{K}^{2}]}+3(\tilde{\gamma}^{jk}\mathcal{K}^{i}{}_{k}+\tilde{\gamma}^{ik}\mathcal{K}^{j}{}_{k})F_{2,[\mathcal{K}^{3}]}\right].
\end{equation}
After a straightforward calculation, we can write down the Lagrangian
density of the extended-MTMG as 
\begin{eqnarray}
\mathcal{L} & = & \frac{\Mpl^{2}}{2}\sqrt{\gamma}N\,[\gamma^{ij}\gamma^{kd}(K_{ik}K_{jd}-K_{ij}K_{kd})+R]\nonumber \\
 & - & \frac{1}{2}m^{2}\Mpl^{2}\sqrt{\gamma}NF_{1}([\mathfrak{K}],[\mathfrak{K}^{2}],[\mathfrak{K}^{3}])-\frac{1}{2}m^{2}\Mpl^{2}\sqrt{\tilde{\gamma}}MF_{2}([\mathcal{K}],[\mathcal{K}^{2}],[\mathcal{K}^{3}])\nonumber \\
 & + & \frac{m^{4}\Mpl^{2}\lambda^{2}M^{2}}{64N}\,\sqrt{\gamma}\gamma_{ik}\gamma_{jd}(2\Theta^{ij}\Theta^{kd}-\Theta^{ik}\Theta^{jd})\nonumber \\
 & + & \lambda\sqrt{\gamma}\left[\mathcal{C}_{\zeta}-\frac{1}{4}\,m^{2}\Mpl^{2}M\,K_{ij}\Theta^{ij}\right]+\sqrt{\gamma}\,(D_{j}\lambda^{i})\mathcal{C}^{j}{}_{i}\,.
\end{eqnarray}
It should be pointed out that the constraints imposed, at the level
of the Lagrangian, impose a non-trivial relation not only on the three-dimensional
metric, but also on the extrinsic curvature. This structure then is
intrinsically different from the Lorentz-breaking massive gravity
theories of \citep{Blas:2009my,Comelli:2014xga}.

The bottom line here is that we have extended MTMG to a more general class of massive gravity theories, which all only possess, at the fully nonlinear level, two tensor-type degrees of freedom on any background. We call the new theory the extended MTMG (eMTMG)~\footnote{We name these models as ``extended'' because they possess a general graviton mass term at the level of the Hamiltonian.}. The original MTMG is a particular case of eMTMG and it can be refound when the functions $F_{1,2}$ reduce to this special form:
\begin{eqnarray}
F_{1}^{{\rm MTMG}} & = & c_{1}\left(\frac{1}{3}\,[\mathfrak{K}^{3}]-\frac{1}{2}[\mathfrak{K}][\mathfrak{K}^{2}]+\frac{1}{6}\,[\mathfrak{K}]^{3}\right)+\frac{1}{2}\,c_{2}\,([\mathfrak{K}]^{2}-[\mathfrak{K}^{2}])+c_{3}\,[\mathfrak{K}]+c_{4}\,,\\
F_{2}^{{\rm MTMG}} & = & c_{1}\,[\mathcal{K}]+\frac{1}{2}\,c_{2}\left([\mathcal{K}]^{2}-[\mathcal{K}^{2}]\right)+c_{3}\left(\frac{1}{3}\,[\mathcal{K}^{3}]-\frac{1}{2}\,[\mathcal{K}][\mathcal{K}^{2}]+\frac{1}{6}\,[\mathcal{K}]^{3}\right),
\end{eqnarray}
as shown in appendix \ref{app:MTMG-subcase}.

\section{Homogeneous and isotropic background}

\label{sec:background}

So far we have extended the original MTMG theory to a much larger
class of theories which is defined out of two free functions $F_{1,2}$
each dependent on three variables. This class of theories is expected
to include a very large set of possibilities in terms of phenomenology.
However, the original motivation to introduce such a class of theories
was, and still is, to cure the problems encountered in the normal branch of MTMG, namely
the presence of a pole in the function $G_{{\rm eff}}/G_{N}$, which would 
in turn lead to an unviable cosmology in a neighborhood
of them\footnote{At the pole itself, at least at linear order, the theory would exit
the regime of validity of a low-energy effective theory description.}. Then it would be interesting to study whether inside the class of
eMTMG theories, it is possible to find a subset which is always
phenomenologically acceptable. By ``always'' we mean for any redshift
and for any background dynamics. This extra dynamical condition might
be too strong, as effectively, one would need only a subset of well-defined
dynamics, however, after imposing it, if such a subset existed, would
provide a ghost-free, instability-free arena, where we
can try to solve today's tensions in cosmology out of a massive graviton.

Hence, let us explore these extended models as to find a good behavior 
for $G_{{\rm eff}}/G_{N}$, the effective gravitational constant for
the density perturbations of a pressure-less fluid on a homogeneous
and isotropic background. For this aim, let us study in this section,
first of all, the background for these theories in the presence of
matter fields. Let us focus then on a spatially flat FLRW background
which is described by 
\begin{equation}
N=N(t)\,,\qquad N^{i}=0\,,\qquad\gamma_{ij}=a(t)^{2}\,\delta_{ij}\,,\qquad\lambda=\lambda(t)\,,\qquad\lambda^{i}=0\,,
\end{equation}
whereas the fiducial sector is given by 
\begin{equation}
M=M(t)\,,\qquad\tilde{\gamma}_{ij}=\tilde{a}(t)^{2}\,\delta_{ij}\,.
\end{equation}

For the matter sector we introduce a perfect fluid (one for each matter
component) modeled by the Schutz-Sorkin action as in~\citep{Schutz77,Brown93,Pookkillath:2019nkn}
\begin{equation}
S_{m}=-\int d^{4}x\sqrt{-g}\,[\rho(n)+J^{\mu}\partial_{\mu}\ell]\,,\qquad n\equiv\sqrt{-J^{\mu}J^{\nu}g_{\mu\nu}}\,,
\end{equation}
for which we can introduce the normalized fluid 4-velocity as $u^{\alpha}=J^{\alpha}/n$,
and $g_{\mu\nu}$ is the four-dimensional physical metric written
in the ADM splitting (\ref{eqn:4dmetric-ADM}). On a spatially flat
FLRW background we have at the level of the background 
\begin{equation}
J^{0}(t)=\frac{\mathcal{J}(t)}{N(t)}\,,\qquad\mathcal{J}(t)=n(t)=\frac{\mathcal{N}_{0}}{a^{3}}\,,
\end{equation}
and the proportionality constant $\mathcal{N}_{0}$ determines the
constant number of fluid particles ($n$ being their number density).
Furthermore, the background equations of motion imply 
\begin{equation}
\ell(t)=-\int^{t}N(t')\rho_{,n}(t')\,dt'\,.
\end{equation}

For the spatially flat FLRW background we find 
\begin{equation}
[\mathfrak{K}^{n}]=3\left(\frac{\tilde{a}}{a}\right)^{n}\,,\qquad[\mathcal{K}^{n}]=3\left(\frac{a}{\tilde{a}}\right)^{n},
\end{equation}
so that the minisuperspace Lagrangian density evaluated on the background
reduces to 
\begin{eqnarray}
\mathcal{L}_{\text{mini}} & = & \frac{3}{2}\left[\dot{a}\left(3F_{2,[\mathcal{K}^{3}]}a^{2}+2F_{2,[\mathcal{K}^{2}]}a\tilde{a}+F_{2,[\mathcal{K}]}\tilde{a}^{2}\right)\,\frac{M}{N}-\dot{\tilde{a}}\left(a^{2}F_{1,[\mathfrak{K}]}+2F_{1,[\mathfrak{K}^{2}]}a\tilde{a}+3F_{1,[\mathfrak{K}^{3}]}\tilde{a}^{2}\right)\right]m^{2}\Mpl^{2}\lambda-\frac{3\Mpl^{2}a\,\dot{a}^{2}}{N}\nonumber \\
 & - & \frac{3M^{2}\left(3F_{2,[\mathcal{K}^{3}]}a^{2}+2F_{2,[\mathcal{K}^{2}]}a\tilde{a}+F_{2,[\mathcal{K}]}\tilde{a}^{2}\right)^{2}m^{4}\Mpl^{2}\lambda^{2}}{16aN}-\frac{m^{2}\Mpl^{2}\tilde{a}^{3}(MF_{2}+NF_{1})}{2}\nonumber \\
 & - & a^{3}\sum_{I}[\mathcal{J}_{I}\dot{\ell}_{I}+N\rho_{I}(\mathcal{J}_{I})]\,.
\end{eqnarray}
On evaluating the Euler-Lagrange equations for the fields $N$, $a$,
$\lambda$, $\ell_{I}$ and $\mathcal{J}_{I}$ we find the equations
of motion for the background.

In order to evaluate the value of $\lambda$ on the background it
turns out to be much simpler to study the Hamilton equations of motion.
Out of the Lagrangian we can find the Hamiltonian in the minisuperspace
via a Legendre transformation as 
\begin{eqnarray}
H_{\text{mini}} & = & \!\left(\frac{p_{a}\left(3F_{2,[\mathcal{K}^{3}]}a^{2}+2F_{2,[\mathcal{K}^{2}]}a\tilde{a}+F_{2,[\mathcal{K}]}\tilde{a}^{2}\right)M}{4a}+\frac{3\Mpl^{2}\dot{\tilde{a}}\left(a^{2}F_{1,[\mathfrak{K}]}+2F_{1,[\mathfrak{K}^{2}]}a\tilde{a}+3F_{1,[\mathfrak{K}^{3}]}\tilde{a}^{2}\right)}{2}\right)m^{2}\lambda\nonumber \\
 & + & \frac{m^{2}\Mpl^{2}\tilde{a}^{3}MF_{2}}{2}+N\left(\frac{F_{1}a^{3}m^{2}\Mpl^{2}}{2}+a^{3}\sum_{I}\rho_{I}(\mathcal{J}_{I})-\frac{p_{a}^{2}}{12a\Mpl^{2}}\right)+\sum_{I}l_{I}p_{\mathcal{J}I}+\sum_{I}\tilde{l}_{I}(\mathcal{J}_{I}a^{3}+p_{\ell I})\,,
\end{eqnarray}
where $p_{a}$, $p_{\mathcal{J}I}$, and $p_{\ell I}$ are the 
momenta conjugate to the variables $a$, $\mathcal{J}_{I}$, and $\ell_{I}$
respectively, whereas $\lambda$, $N$, $l_{I}$ and $\tilde{l}_{I}$
are all Lagrange multipliers which set constraints. One such constraint
is then 
\begin{equation}
C_{0}=\frac{p_{a}\left(3F_{2,[\mathcal{K}^{3}]}a^{2}+2F_{2,[\mathcal{K}^{2}]}a\tilde{a}+F_{2,[\mathcal{K}]}\tilde{a}^{2}\right)M}{4a}+\frac{3\Mpl^{2}\dot{\tilde{a}}\left(a^{2}F_{1,[\mathfrak{K}]}+2F_{1,[\mathfrak{K}^{2}]}a\tilde{a}+3F_{1,[\mathfrak{K}^{3}]}\tilde{a}^{2}\right)}{2}\approx0\,,
\end{equation}
whereas the Hamiltonian constraint can be written as 
\begin{equation}
R_{0}=\frac{F_{1}a^{3}m^{2}\Mpl^{2}}{2}+a^{3}\sum_{I}\rho_{I}(\mathcal{J}_{I})-\frac{p_{a}^{2}}{12a\Mpl^{2}}\approx0\,.
\end{equation}
Let us now impose that the time derivative of the constraints should
vanish on the constraint surface. For example we have 
\begin{equation}
\dot{p}_{\mathcal{J}I}=\{p_{\mathcal{J}I},H_{\text{mini}}\}=-a^{3}(N\rho_{I,\mathcal{J}}+\tilde{l}_{I})\approx0\,,\qquad{\rm or}\qquad\tilde{l}_{I}\approx-N\rho_{I,\mathcal{J}}\,,
\end{equation}
which sets all the $\tilde{l}_{I}$'s Lagrange multipliers in the
matter sectors. Furthermore we have 
\begin{equation}
\{\mathcal{J}_{I}a^{3}+p_{\ell I},H_{\text{mini}}\}=3\mathcal{J}_{I}\,a^{2}\left(\frac{M\,m^{2}\left(3F_{2,[\mathcal{K}^{3}]}a^{2}+2F_{2,[\mathcal{K}^{2}]}a\tilde{a}+F_{2,[\mathcal{K}]}\tilde{a}^{2}\right)\lambda}{4a}-\frac{p_{a}N}{6\Mpl^{2}a}\right)+a^{3}l_{I}\approx0\,,
\end{equation}
which can be used in order to set the Lagrange multipliers $l_{I}$'s.
Also we can find 
\begin{equation}
\dot{R}_{0}\equiv\{R_{0},H_{\text{mini}}\}+\frac{\partial R_{0}}{\partial\tilde{a}}\,\dot{\tilde{a}}\approx0\,,
\end{equation}
which combined with $C_{0}$, gives 
\begin{equation}
\dot{R}_{0}-C_{0}\approx f\,\lambda\approx0\,,
\end{equation}
where $f$ is a quantity which in general does not vanish, unless
some fine-tuned dynamics are considered. This equation then determines
the Lagrange multiplier $\lambda$, without adding any new constraint,
and it finally leads to the conclusion that on the constraint surface,
that is on the background, we need to impose $\lambda(t)=0$.

With $\lambda(t)=0$, the independent background equations of motion
greatly simplify and reduce to 
\begin{eqnarray}
3\Mpl^{2}H^{2} & = & \sum_{I}\rho_{I}+\frac{1}{2}\,\Mpl^{2}\,m^{2}\,F_{1}\,,\\
H\,\frac{M}{N}\left[\left(\frac{\tilde{a}}{a}\right)^{2}F_{2,[\mathcal{K}]}+2\left(\frac{\tilde{a}}{a}\right)F_{2,[\mathcal{K}^{2}]}+3F_{2,[\mathcal{K}^{3}]}\right] & = & \frac{\dot{\tilde{a}}}{N\tilde{a}}\,\frac{\tilde{a}}{a}\left[F_{1,[\mathfrak{K}]}+2\left(\frac{\tilde{a}}{a}\right)F_{1,[\mathfrak{K}^{2}]}+3\left(\frac{\tilde{a}}{a}\right)^{2}F_{1,[\mathfrak{K}^{3}]}\right],\label{eq:constr_lb_1}\\
\frac{\dot{\rho}_{I}}{N} & = & -3\,H\,(\rho_{I}+P_{I})\,,
\end{eqnarray}
where $H\equiv\dot{a}/(Na)$ is the Hubble expansion rate for the
physical metric. Let us then define 
\begin{equation}
X\equiv\frac{\tilde{a}}{a}\,,
\end{equation}
and suppose that $X>0$, during the whole evolution of the universe
in the regime of interest. The constraint equation, Eq.\ (\ref{eq:constr_lb_1}),
can be rewritten as 
\begin{equation}
H\,\frac{M}{N}\left(X^{2}F_{2,[\mathcal{K}]}+2XF_{2,[\mathcal{K}^{2}]}+3F_{2,[\mathcal{K}^{3}]}\right)=\biggl(\frac{\dot{X}}{N}+HX\biggr)\left(F_{1,[\mathfrak{K}]}+2XF_{1,[\mathfrak{K}^{2}]}+3X^{2}F_{1,[\mathfrak{K}^{3}]}\right).\label{eq:constr_FLRW}
\end{equation}

Unlike the original MTMG, the space of solutions for Eq.\ (\ref{eq:constr_FLRW})
is not in general separated into two branches, the so called self-accelerating
and normal branches. The special case in which such separation occurs
is briefly studied in Appendix~\ref{app:selfacce}. On the other
hand, in the rest of the present paper we consider the general case,
that is the single universal branch, defined by Eq.\ (\ref{eq:constr_FLRW}),
for all the eMTMG models.

\section{Linear perturbations}

\label{sec:linearperturbations}

In this section we study linear perturbations around the spatially
flat FLRW background introduced in the previous section.

\subsection{Gravitational waves}

\label{subsec:gw}

Let us now consider the tensor perturbations for the physical metric,
namely
\begin{equation}
N=N(t)\,,\qquad N^{i}=0\,,\qquad\gamma_{ij}=a^{2}\left[\delta_{ij}+\sum_{\lambda={+},{\times}}\epsilon_{ij}^{\lambda}\,h_{\lambda}\right]\,,
\end{equation}
where the two symmetric polarization tensors satisfy both the transverse
and traceless conditions $\epsilon_{ij}^{\lambda}\delta^{jl}\partial_{l}h_{\lambda}=0$,
$\delta^{ij}\epsilon_{ij}^{\lambda}=0$, and the chosen normalizations
$\epsilon_{ij}^{{+}}\epsilon_{lm}^{{+}}\delta^{il}\delta^{jm}=1=\epsilon_{ij}^{{\times}}\epsilon_{lm}^{{\times}}\delta^{il}\delta^{jm}$,
together with $\epsilon_{ij}^{{+}}\epsilon_{lm}^{{\times}}\delta^{il}\delta^{jm}=0$.
After expanding the Lagrangian at the second order in the tensor perturbations,
we obtain the quadratic action describing their dynamics as 
\begin{equation}
S=\frac{\Mpl^{2}}{8}\sum_{\lambda={+},{\times}}\int d^{4}x\,N\,a^{3}\left[\left(\frac{\dot{h}_{\lambda}}{N}\right)^{2}-\frac{(\partial h_{\lambda})^{2}}{a^{2}}-\mu^{2}\,h_{\lambda}^{2}\right],
\end{equation}
where 
\begin{equation}
\mu^{2}=\frac{1}{2}\,m^{2}X\left[r\left(X^{2}F_{2,[\mathcal{K}]}+4XF_{2,[\mathcal{K}^{2}]}+9F_{2,[\mathcal{K}^{3}]}\right)+F_{1,[\mathfrak{K}]}+4XF_{1,[\mathfrak{K}^{2}]}+9X^{2}F_{1,[\mathfrak{K}^{3}]}\right],\label{eq:mass_gw}
\end{equation}
and we have defined for later convenience also the quantity 
\begin{equation}
r\equiv\frac{1}{X}\frac{M}{N}\,.
\end{equation}
Therefore these models do introduce a non-trivial mass for the tensor
modes, however the speed of propagation, for high-$k$ modes, i.e.\ at
energies for which the graviton becomes ultra-relativistic, will still
be equal to unity. Furthermore, the graviton mass not only does not
vanish in general, but also it is changing with time. For this reason,
we will also demand that well-behaved subset of eMTMG models
would also satisfy the condition of a non-negative $\mu^{2}$ for
any dynamics of the background.

\subsection{Effective gravitational constant}

\label{subsec:geff}

In this section we consider instead the scalar perturbations and expand
the action for the theory in the presence of matter perfect fluids
up to second order and remove at the level of the action all the auxiliary
fields to find the quadratic action for the field $\delta\rho/\rho$, which we shall define
below.

First of all we will explicitly write down all the perturbation variables,
both in the gravity and in the matter sector. We introduce scalar
perturbations for the physical metric in the following way:
\begin{eqnarray}
N & = & N(t)\,(1+\alpha)\,,\\
N_{i} & = & N(t)\,\partial_{i}\chi\,,\\
\gamma_{ij} & = & a(t)^{2}\,\delta_{ij}\,(1+2\zeta)+2\partial_{ij}E\,.
\end{eqnarray}
Since we have fixed from the beginning the unitary gauge, we cannot
impose any gauge condition on the perturbation variables. We also
need to introduce perturbations for the following eMTMG
variables 
\begin{equation}
\lambda=\delta\lambda\,,\qquad\lambda^{i}=\frac{1}{a^{2}}\,\delta^{ij}\,\partial_{j}\delta\lambda_{V}\,.
\end{equation}
As for the matter sectors we proceed instead as follows. First of
all we make the following split 
\begin{eqnarray}
J^{0} & = & \frac{\mathcal{J}(t)}{N(t)}\,(1+\delta J)\,,\\
J^{i} & = & \frac{1}{a^{2}}\,\delta^{ij}\partial_{j}\delta J_{V}\,,\\
\ell & = & \ell(t)+\delta\ell\,.
\end{eqnarray}
For each matter component we consider matter field redefinitions as
follows. We first define the fluid perturbation scalar velocity $v$
as 
\begin{equation}
u_{i}=g_{i\mu}\,u^{\mu}=\partial_{i}v\,,
\end{equation}
which leads to the field redefinition 
\begin{equation}
\delta J_{V}=n(t)\,(v-\chi)\,.
\end{equation}
Expanding the action at second order in the perturbation variables,
finding the equation of motion for $v$ and solving it for $\delta\ell$
gives 
\begin{equation}
\delta\ell=\rho_{,n}\,v\,,
\end{equation}
which can be used in order to integrate out the field $\delta\ell$.
Also we can perform a field redefinition as follows 
\begin{equation}
\delta J=\frac{\rho}{n\rho_{,n}}\,\frac{\delta\rho}{\rho}-\alpha\,,\qquad{\rm where}\qquad\frac{\delta\rho}{\rho}\equiv\frac{\rho}{\rho(t)}-1\,.
\end{equation}

As for now we have an action for the perturbation which is a function
of the following variables: $\alpha,\chi,\zeta,E$ and $\delta\lambda,\delta\lambda_{V}$
in the metric sector, together with $\delta\rho/\rho$ and $v$ for
each matter-fluid component. We can find equations of motion for each
of these perturbation variables, and we label them, e.g\ as $E_{\chi}$
(which vanish, i.e.\ $E_{\chi}=0$, and the subscript shows the variables
for which the equation of motion is derived, $\chi$ in this example).
In the following, although not necessary, we will use time-reparametrization
as to set $N(t)=a(t)$. Since we want to match the phenomenology with
observations we will also make the following field redefinitions which
link $\alpha$ and $\zeta$ to the gauge-invariant definitions of
the Bardeen potentials $\psi$ and $\phi$: 
\begin{eqnarray}
\alpha & = & \psi-\frac{1}{a}\,\dot{\chi}+\frac{1}{a}\,\frac{d}{dt}\!\left[a\frac{d}{dt}\!\left(\frac{E}{a^{2}}\right)\right],\label{eq:alpha_GI}\\
\zeta & = & -\phi-H\,\chi+a\,H\,\frac{d}{dt}\!\left(\frac{E}{a^{2}}\right),\\
\frac{\delta\rho}{\rho} & = & \delta-\frac{\dot{\rho}}{a\rho}\,\chi+\frac{\dot{\rho}}{\rho}\,\frac{d}{dt}\!\left(\frac{E}{a^{2}}\right)\,,
\end{eqnarray}
whereas the last equation introduces $\delta$ as the gauge invariant
longitudinal matter perturbation. Finally we also make the field redefinition
\begin{equation}
v=-\frac{a}{k^{2}}\,\theta+\chi-a\,\frac{d}{dt}\!\left(\frac{E}{a^{2}}\right),\label{eq:v_GI}
\end{equation}
where $\theta$ is another gauge invariant variable related to the
scalar fluid velocity. We can also introduce, at the level of perturbation,
a shear term for each matter component as done in \citep{VCDM:solvingH0}.

So far, the equation of state of the perfect fluid is general and in
the next subsection we shall further consider equations of motion for
this general system. In the rest of this subsection, on the other
hand, we restrict our considerations to the case of a single perfect
fluid to compute its sound speed and in the case of dust, the
effective gravitational constant.

The first non-trivial feature of the models consists in the constraint
equation set by the field $\delta\lambda_{V}$. In fact we find that
it can be written as
\begin{eqnarray}
\zeta & \propto & \Xi_{1}\,\frac{k^{2}}{a^{2}}E\,.\\
\Xi_{1} & \equiv & F_{2,[\mathcal{K}][\mathcal{K}]}+\frac{4}{X}\,F_{2,[\mathcal{K}][\mathcal{K}^{2}]}+\frac{6}{X^{2}}\,F_{2,[\mathcal{K}][\mathcal{K}^{3}]}+\frac{4}{X^{2}}\,F_{2,[\mathcal{K}^{2}][\mathcal{K}^{2}]}+\frac{12}{X^{3}}\,F_{2,[\mathcal{K}^{2}][\mathcal{K}^{3}]}\nonumber \\
 & + & \frac{9}{X^{4}}\,F_{2,[\mathcal{K}^{3}][\mathcal{K}^{3}]}+2\,F_{2,[\mathcal{K}^{2}]}+\frac{6}{X}\,F_{2,[\mathcal{K}^{3}]}\,.
\end{eqnarray}
Therefore, the quantity $\Xi_{1}$ discriminates the behavior of the
theory in the high-$k$ regime, as, in general, the phenomenology
of the theory will be different for the eMTMG models depending
on whether $\Xi_{1}$ is zero (or negligible) or not. The mirror quantity
$\Xi_{2}$ for the function $F_{1}$, turns out to have also a strong
influence on the phenomenology of the theory, as we will see later
on

\begin{eqnarray}
\Xi_{2} & = & F_{1,[\mathfrak{K}][\mathfrak{K}]}+4X\,F_{1,[\mathfrak{K}][\mathfrak{K}^{2}]}+6X^{2}F_{1,[\mathfrak{K}][\mathfrak{K}^{3}]}+4X^{2}F_{1,[\mathfrak{K}^{2}][\mathfrak{K}^{2}]}+12X^{3}F_{1,[\mathfrak{K}^{2}][\mathfrak{K}^{3}]}\nonumber \\
 &  & +9X^{4}F_{1,[\mathfrak{K}^{3}][\mathfrak{K}^{3}]}+2F_{1,[\mathfrak{K}^{2}]}+6X\,F_{1,[\mathfrak{K}^{3}]}\,.
\end{eqnarray}
Indeed, one can proceed to remove all the auxiliary fields except
for the field $\delta\rho/\rho$, which, in the case of a single fluid,
has the following schematic quadratic Lagrangian density
\begin{equation}
\mathcal{L}_{\delta\rho}=A(k^{2},t)\left[\frac{1}{N}\,\frac{\partial}{\partial t}\!\left(\frac{\delta\rho}{\rho}\right)\right]^{2}+B(k^{2},t)\left(\frac{\delta\rho}{\rho}\right)^{2}.\label{eq:lag_dens}
\end{equation}
In the high-$k$ regimes, we find that the no-ghost condition is always
verified, since
\begin{eqnarray}
A & = & \frac{1}{2}\,Na^{3}\,\frac{a^{2}}{k^{2}}\,\frac{\rho^{2}}{n\rho_{,n}}+\mathcal{O}(a^{4}/k^{4})\,,
\end{eqnarray}
which is always positive, provided that $n\rho_{,n}=\rho+P>0$. As
for the $B$ term, we need to distinguish among possibilities.
\begin{itemize}
\item Case for which $\Xi_{1}\neq0$, and in this case we have
\begin{equation}
B=-\frac{Na^{3}}{2}\,\frac{\rho_{,nn}\rho^{2}}{\rho_{,n}^{2}}+\mathcal{O}(a^{2}/k^{2})\,,\qquad{\rm or}\qquad c_{s}^{2}=\frac{n\rho_{,nn}}{\rho_{,n}}\,,
\end{equation}
giving the standard results for the propagation of perturbations in
a fluid.
\item Case for which $\Xi_{1}=0$, or very negligible namely $\Xi_{1}k^{2}/(a^{2}H^{2})\ll1$,
and in this case we find instead
\begin{equation}
B=-\frac{Na^{3}}{2}\left[\frac{\rho_{,nn}\rho^{2}}{\rho_{,n}^{2}}+\Xi_{2}\,B_{2}(t)\,\frac{m^{2}\rho^{2}}{\Mpl^{2}H^{4}}\right]+\mathcal{O}(a^{2}/k^{2})\,,\qquad{\rm or}\qquad c_{s}^{2}=\frac{n\rho_{,nn}}{\rho_{,n}}+\Xi_{2}\,B_{2}\,\frac{m^{2}}{H^{2}}\,\frac{\rho+P}{\Mpl^{2}H^{2}}\,,
\end{equation}
which leads to a non-trivial propagation speed unless also $\Xi_{2}=0$
(or, as mentioned above, very negligible). Indeed the case $\Xi_{1}=0=\Xi_{2}$,
is the one we are going to focus on in the following sections\footnote{Since $B_{2}\propto\left(X^{2}F_{2,[\mathcal{K}]}+2XF_{2,[\mathcal{K}^{2}]}+3F_{2,[\mathcal{K}^{3}]}\right)^{2}\left(F_{1,[\mathfrak{K}]}+2XF_{1,[\mathfrak{K}^{2}]}+3X^{2}F_{1,[\mathfrak{K}^{3}]}\right)^{2}$,
the case $B_{2}=0$ corresponds to the self-accelerating case, which
will be discussed in Appendix \ref{app:selfacce}.}. Nonetheless, as long as $\text{\ensuremath{\Xi_{2}}}$ (or $B_{2}$)
does not vanish, the speed of propagation for any matter fluid will
get modified. This nontrivial property is shared by another minimal
theory of gravity introduced and studied in \citep{Mukohyama:2019unx,MMTG_Planck}.
In particular, a pressureless fluid will acquire a nontrivial contribution.
\item Case for which $\Xi_{1}=0=\Xi_{2}$ and $\rho_{,nn}=0=c_{s}^{2}$,
where the last equation of state corresponds to choosing a pressureless
fluid as matter field. In this case the Lagrangian density of Eq.\ (\ref{eq:lag_dens})
for the energy-density perturbations reduce to
\begin{equation}
\mathcal{L}_{{\rm dust}}=\frac{1}{2}\,Na^{3}\,\frac{a^{2}}{k^{2}}\,\rho\left\{ \left[\frac{1}{N}\frac{\partial}{\partial t}\!\left(\frac{\delta\rho}{\rho}\right)\right]^{2}+4\pi G_{{\rm eff}}\,\rho\left(\frac{\delta\rho}{\rho}\right)^{2}\right\} ,
\end{equation}
out of which one can deduce the expression for $G_{{\rm eff}}/G_{N}$,
whose value (which is not unity, in general) is explicitly written
in Appendix \ref{app:Geff_GN}.
\end{itemize}

\subsection{Equations of motion for scalar perturbations}

\label{subsec:eom-scalarperturbation}

In this subsection, instead of considering a Lagrangian approach,
we consider an equivalent approach, based on studying the equations
of motion for scalar perturbations in the presence of matter fields
modeled by perfect fluids with general equations of state. In this
case though, we generalize the previous results to the case of an
arbitrary number of matter fields.

In particular, in terms of the gauge invariant variables introduced
in Eqs.\ (\ref{eq:alpha_GI})--(\ref{eq:v_GI}), the matter equations
of motion for each matter component are the same as in General Relativity,
namely 
\begin{eqnarray}
\dot{\delta}_{I} & = & 3a(w_{I}-c_{sI}^{2})H\delta_{I}-(1+w_{I})\,(\theta_{I}-3\dot{\phi})\,,\label{eq:dot_d_gen}\\
\dot{\theta}_{I} & = & aH(3c_{sI}^{2}-1)\,\theta_{I}+k^{2}\psi+\frac{c_{sI}^{2}k^{2}}{1+w_{I}}\,\delta_{I}-k^{2}\sigma_{I}\,,\label{eq:dot_th_gen}
\end{eqnarray}
where $w_{I}\equiv P_{I}/\rho_{I}$, and $c_{sI}^{2}=\dot{p}_{I}/\dot{\rho}_{I}=\left(\frac{\partial p_{I}}{\partial\rho_{I}}\right)_{s}$
is the speed of propagation for each matter species. Here the subscript
$I$ runs over all the standard matter components we consider. The
fact that in the matter sector we refind the same equations of GR
is not surprising, as the Lagrangians of matter fields do satisfy general
covariance.

We can proceed by solving $E_{\delta\lambda_{V}}$ for $\chi$, $E_{\alpha}$
for $E$ and $E_{\chi}$ for $\delta\lambda$. Now the equation of
motion $E_{E}$ can be written as 
\begin{equation}
E_{E}=S_{1}\dot{\phi}+S_{2}\phi+S_{3}\psi+\sum_{I}S_{4,I}\dot{\delta}_{I}+\sum_{I}S_{6,I}\delta_{I}+\sum_{I}S_{8,I}\dot{\theta}_{I}+\sum_{I}S_{10,I}\theta_{I}+S_{12}\delta\lambda_{V}=0\,,\label{eq:dyn1_aux}
\end{equation}
where the $S$'s coefficients are functions of $k$ and time\footnote{The explicit form of the coefficients $S_{1,\dots,12}$ are in general
quite lengthy, and not strongly illuminating. We will instead write
the explicit expressions of the observables when needed.}. A linear combination of $E_{E}$ and $E_{\zeta}$ leads instead
to 
\begin{equation}
E_{E\zeta}=T_{1}\delta\lambda_{V}+T_{2}\phi+T_{3}\psi+\sum_{I}T_{4,I}\delta_{I}+\sum_{I}T_{6,I}\theta_{I}+\sum_{I}T_{6,I}\sigma_{I}=0\,,\label{eq:E_and_zeta}
\end{equation}
which can be used to define $\delta\lambda_{V}$ in terms of the other
variables. Finally the equation $E_{\delta\lambda}$ leads to 
\begin{equation}
E_{\delta\lambda}=U_{1}\phi+\sum_{I}U_{2,I}\delta_{I}+\sum_{I}U_{4,I}\theta_{I}=0\,.\label{eq:dyn_0-1}
\end{equation}
Here, $T$'s and $U$'s are, once more, coefficients which depend
on $k$ and time $t$. On considering the time derivative of Eq.\ (\ref{eq:dyn_0-1}),
namely $\dot{E}_{\delta\lambda}$, and replacing $\dot{\phi}$, $\dot{\delta}_{I}$
and $\dot{\theta}_{I}$ with those given by Eqs.\ (\ref{eq:dyn1_aux}), (\ref{eq:dot_d_gen})
and (\ref{eq:dot_th_gen}) respectively, we arrive at the so called
``shear equation'' for this theory. This approach then leads to
a structure of the equations of motion which is analogue to the standard
approach in General Relativity. This is a consequence of the fact
that these theories do not add any new degree of freedom, i.e.\ no
new dynamical equation is necessary to determine new fields (which
do not exist in the first place), but still they change the equations
of motion for the linear perturbation e.g.\ by the fact that the coefficients
$S_{1,\dots,12}$ differ from the ones in GR. So after all, a different
phenomenology is expected to take place in general, although the number
of degrees of freedom has not been changed.

\section{Phenomenological criteria}

\label{sec:criteria}

So far the approach was totally general, and it can be applied to 
general functions $F_{1,2}$ and arbitrary matter components.
On the other hand, in order to give predictions on the graviton mass
from observational data, it is ideal to have a well-motivated subclass
of models with a finite number of parameters. For this reason, in
this section we shall make a list of phenomenologically motivated
criteria to be imposed on the theory.

\subsection{$c_{s}^{2}=0$ at all times and $G_{{\rm eff}}/G_{N}\to1$ at early
times for pressureless fluids}

\label{subsec:criteronGefftoGN@earlytime}

In subsection~\ref{subsec:geff} we have computed the squared sound
speed $c_{s}^{2}$ and the effective gravitational constant $G_{{\rm eff}}$
for scalar perturbations in the presence of a pressureless fluid.
Based on these considerations and the result of subsection~\ref{subsec:geff},
we can divide the theories into the following three categories. 
\begin{enumerate}
\item The case $\Xi_{1}\neq0$. In this case, from the equation of motion
for $\lambda^{i}$, we find $\zeta\propto E$. Furthermore, after
removing all the auxiliary fields, we find that $G_{{\rm eff}}/G_{N}=1+\mathcal{O}([k/(aH)]^{-2})$
independently of the model. Therefore, for this class of models, the
background dynamics will be modified, but the behavior of the perturbations,
in the short scales regime, will not. This could be an interesting
possibility, but since, we are looking for theories which can address
a gravitational interaction weaker than General Relativity, we will
not discuss this model further in this paper, but we will consider
it as a subject of investigation for a future project. 
\item The case $\Xi_{1}=0$, but $\Xi_{2}\neq0$. In this case, $\zeta=0$,
and the matter perturbations acquire a non-zero speed of propagation,
namely $c_{s}^{2}\neq0$. This case will not be discussed further
as strongly constrained from a phenomenological point of view. 
\item The case $\Xi_{1}=0=\Xi_{2}$. In this case, $\zeta=0$, $G_{{\rm eff}}/G_{N}=A(t)/Z(t)+\mathcal{O}([k/(aH)]^{-2})$,
where $A(t)$ and $Z(t)$ are expressions which only depend on time.
In general $G_{{\rm eff}}/G_{N}\neq1$, but 
\begin{equation}
\lim_{m/H\to0}\frac{G_{{\rm eff}}}{G_{N}}=1\,,
\end{equation}
which leads to the standard evolution for matter perturbations at early
times. 
\end{enumerate}
Both $c_{s}^{2}$ and $G_{{\rm eff}}$ are important for the formation
of the large-scale structure in the universe. In GR, $c_{s}^{2}$
for a pressureless fluid vanishes and this is consistent with observations.
If $c_{s}^{2}$ is negative or too large then the prediction of the
theory would contradict with observations. Although a positive and
sufficiently small $c_{s}^{2}$ can in principle be consistent, for
simplicity we restrict our considerations to the case with $c_{s}^{2}=0$
for the eMTMG. In this case the effective gravitational constant
controls the behavior of scalar perturbations. Motivated by the fact
that the standard $\Lambda$CDM in GR fits the Planck data very well,
we demand that $G_{{\rm eff}}/G_{N}\to1$ at early times. On the other
hand, at late times we would like to have deviation of $G_{{\rm eff}}/G_{N}$
from unity so that we may hope to address tensions in cosmology. For
these reasons, in the rest of the present paper we study the third
case above, that is the case for which 
\begin{equation}
\Xi_{1}=0=\Xi_{2}\,.
\end{equation}

\subsection{Finite $G_{{\rm eff}}/G_{N}$ for $^{\forall}X(t)$}

One of the main motivation in this paper is to have models for which
$G_{{\rm eff}}/G_{N}$ never blows up during a general evolution of
$X(t)$ and $r(t)$, keeping the theory safely in the regime of validity 
of the effective field theory. This implies that, on considering the quantity
$Z(t)$, we need to impose that it never vanishes for any dynamics
of $X(t)$ and $r(t)$, assuming them to be all positive quantities.

If the expression for $G_{{\rm eff}}/G_{N}$ has poles for some real
value of $X=X_{\infty}$, we could just make sure that during the
dynamics these values of $X_{\infty}$ should not be reached. However,
this might not be possible to predict in general. For example in MTMG,
this pole corresponds to a time for which $H^{2}=H_{\infty}^{2}=\mu^{2}/2$,
where $\mu$ is the mass of the graviton. This means that for MTMG
we had to have that $H_{\infty}^{2}<H_{0}^{2}$. However, well before
reaching this pole, the phenomenology of the theory was leading to
inconsistencies, see e.g.\ \citep{DeFelice:2016ufg,Hagala:2020eax}. Therefore,
just avoiding the poles may not be enough to lead to a viable phenomenology.
Therefore, we want to find, if possible a subset of theories which,
under this point of view, are always consistent.

\subsection{Positive $\mu^{2}$ for $^{\forall}X(t)$}

Furthermore, we also impose that the squared mass $\mu^{2}$ of the
tensor modes remains finite and positive during a general evolution
of $X(t)$ and $r(t)$.

As previously mentioned, in MTMG, a pole for $G_{{\rm eff}}/G_{N}$
was reached when $H^{2}=H_{\infty}^{2}=\mu^{2}/2$. Evidently this
pole can be removed provided that we impose $\mu^{2}<0$, that is
when the graviton has a negative mass squared. What would this mean?
In a cosmological scenario, this would make tensor modes unstable.
But this instability would be reached when the energy of the graviton
itself, $E=\sqrt{k_{{\rm phys}}^{2}-|\mu^{2}|}$ is comparable to
$\sqrt{|\mu^{2}|}$, and as such astrophysically produced gravitons
will not show this instability. Furthermore, such an instability would
have a typical time-of-instability of order of $1/\sqrt{|\mu^{2}|}\simeq H_{0}^{-1}$,
which would become evident only in the future.

Nonetheless, in this paper we assume the tensor modes to be non-tachyonic.
In this case we have to impose that during the whole history of the
universe $\mu^{2}\geq0$. This condition is, in general, independent
of the absence of poles in $G_{{\rm eff}}/G_{N}$, so we expect that they can be
imposed simultaneously.

\subsection{Finite ISW effect}\label{subssec:ISW_finite}

There are other observables which still strongly influence the behaviour
of late time cosmology. In particular, here we consider the following
combination, whose time derivative affects the ISW effect, namely
\begin{equation}
\psi_{{\rm ISW}}\equiv\phi+\psi\,.
\end{equation}
We demand that $\psi_{{\rm ISW}}$ remains finite during a general
evolution of $X(t)$ and $r(t)$.

In fact, the correlation between ISW and galaxy perturbations usually
sets strong constraints for modified gravity models, in particular
an anticorrelation signal is ruled out. For example in MTMG, it was
shown that the ISW-galaxy correlation effect was one of the most stringent
bound the theory had to pass, see e.g.\ \citep{DeFelice:2016ufg,MTMG:isw,DeFelice:2020cpt,DeFelice:2021trp}.
As already stated above, in order not to have strong constraints coming
from this observable, we demand the finiteness of $\psi_{{\rm ISW}}$
and its time derivative during the whole dynamics of the universe.
Doing so in principle should add new constraints, and as such further
reduce the possibilities for the model to exist. However, as we shall
see later on, at least for the case under study the finiteness of
$G_{{\rm eff}}/G_{N}$ turns out to be a sufficient condition for
the finiteness of the ISW observable.

\section{Concrete realization based on polynomial ansatz for $F_{1,2}$}

\label{sec:polynomialansatz}

In principle it should be possible to find the subclass consisting
of all theories that satisfy the phenomenological criteria summarized
in the previous section. However, the analysis and the result are
expected to be rather complicated (if possible in practice). In this
section we therefore consider a simple ansatz for the functions $F_{1,2}$
and then impose the phenomenological criteria step by step.

Considering the fact that the original MTMG has polynomial expressions
for $F_{1,2}$ in terms of their variables, we therefore restrict
our considerations to the case where $F_{1,2}$ are polynomials of
their arguments. Furthermore, for simplicity we truncate the polynomials
at the six order in $\mathcal{K}^{i}{}_{j}$ and $\mathfrak{K}^{i}{}_{j}$,
respectively, as 
\begin{eqnarray}
F_{1}(A,B,C) & = & a_{3,111111}A^{6}+a_{3,11111}A^{5}+a_{3,1111}A^{4}+a_{3,111}A^{3}+a_{3,112}A^{2}B+a_{3,222}B^{3}+a_{2,11}A^{2}+2a_{2,12}AB\nonumber \\
 & + & a_{3,122}AB^{2}+2a_{2,13}AC+a_{2,22}B^{2}+2a_{2,23}BC+a_{2,33}C^{2}+a_{1,1}A+a_{1,2}B+a_{1,3}C\nonumber \\
 & + & g_{1}A^{4}B+g_{2}A^{2}B^{2}+g_{3}A^{3}B+g_{4}A^{2}C+g_{5}A^{3}C+g_{6}ABC+c_{4}\,,\\
F_{2}(A,B,C) & = & b_{3,111111}A^{6}+b_{3,11111}A^{5}+b_{3,1111}A^{4}+b_{3,111}A^{3}+b_{3,112}A^{2}B+b_{3,222}B^{3}+b_{2,11}A^{2}+2b_{2,12}AB\nonumber \\
 & + & b_{3,122}AB^{2}+2b_{2,13}AC+b_{2,22}B^{2}+2b_{2,23}BC+b_{2,33}C^{2}+b_{1,1}A+b_{1,2}B+b_{1,3}C\nonumber \\
 & + & h_{1}A^{4}B+h_{2}A^{2}B^{2}+h_{3}A^{3}B+h_{4}A^{2}C+h_{5}A^{3}C+h_{6}ABC\,.
\end{eqnarray}
Essentially, as already stated above, the polynomials have been chosen so as to be linear combinations of terms in the form $A^{b_{1}}B^{b_{2}}C^{b_{3}}$, with the conditions that
$b_{i}\in\mathbb{N}^{+}$ and $b_{3}\leq2$. Then the other powers,
$b_{1,2}$, have been chosen so that, on FLRW, once written as polynomials
in $X$, which can be grouped according to equal powers of $X$. For
example, the variable $C$ in $F_{1}$ will lead to a term $X^{3}$,
$B$ to $X^{2}$ and $A$ to $X$. Therefore, for instance, we allow
terms in $C^{2},A^{3}C,ABC,B^{3},A^{2}B^{2},A^{4}B,A^{6}$ which all
lead to a term proportional to $X^{6}$, etc. Although this toy model
is just meant to be a proof-of-existence case, we will find that the
models satisfying the properties we are looking for, they all behave
in the same way on the FLRW background, so that we believe the model
can catch general properties of the extended minimal models of gravity.
Indeed, as we shall see later on, further simplified sub cases which are 
still general enough in their dynamics will be found.

In the following we shall impose all phenomenological criteria considered
in the previous section step by step.

\subsection{$c_{s}^{2}=0$ at all times and $G_{{\rm eff}}/G_{N}\to1$ at early
times}

Let us now impose the conditions on the parameters so that $c_{s}^{2}=0$
at all times, that $G_{{\rm eff}}/G_{N}\to1$ at early times and yet
that $G_{{\rm eff}}/G_{N}$ exhibits interesting deviation from unity
at late times. As discussed in subsection~\ref{subsec:criteronGefftoGN@earlytime},
this amounts to requiring that, for any $X$, we have $\Xi_{1}=0=\Xi_{2}$.
This will select the models belonging to the third case mentioned
above. This conditions fixes some constant parameters to satisfy the
following relations 
\begin{eqnarray}
a_{3,111111} & = & -\frac{7}{135}\,a_{3,222}-\frac{1}{45}\,a_{2,33}\,,\\
a_{3,11111} & = & -\frac{2}{15}\,a_{2,23}-\frac{7}{45}\,a_{3,122}\,,\\
a_{3,1111} & = & -\frac{4}{9}\,a_{2,13}-\frac{5}{27}\,a_{2,22}-\frac{4}{9}\,a_{3,112}\,,\\
a_{3,111} & = & -\frac{1}{3}\,a_{1,3}-\frac{10}{9}\,a_{2,12}\,,\\
a_{2,11} & = & -a_{1,2}\,,
\end{eqnarray}
with analogue relations holding for the $b$'s coefficients.

\subsection{Finite $G_{{\rm eff}}/G_{N}$ for $^{\forall}X(t)$}

We now require that $Z(t)$ never vanishes for any positive $X(t)$
and $r(t)$. Later we shall also impose the positivity of $\mu^{2}(t)$.
Therefore, in this subsection we also assume that $\mu^{2}(t)$ is
also positive.

In order to simplify the expression for $Z(t)$, we first replace
$\dot{X}$ on using the background constraint, Eq.\ (\ref{eq:constr_FLRW}).
We also replace $\dot{a}$ in terms of $H$, $M$ in terms of $r$,
$N$, $X$, and say $b_{3,222}$ in terms of $\mu^{2}$, by inverting
Eq.\ (\ref{eq:mass_gw}). Then we find that $Z\propto Z_{1}(t)^{2}Z_{2}(t)^{2}Z_{3}(t)^{2}Z_{4}(t)^{2}$
(where the proportionality factor is positive definite, by assumptions,
being a product of powers of $H$, $r$, $X$, $a$, $N$) and $Z_{I}$
($I=\{1,\dots,4\}$) are instead polynomial in powers of $X$, $r$
and $\mu^{2}$. We conservatively impose that each of the coefficients
of such polynomials have the same sign, so that each polynomial $Z_{I}$
would never vanish. The expressions $Z_{3,4}$ only set constraints
on the $a$'s parameters, whereas $Z_{1,2}$ also constrain the $b$'s
parameters. On considering only $Z_{2,3,4}$ ($Z_{1}$ being the most
complicated expression) then we find the following constraints need
to be satisfied 
\begin{eqnarray}
a_{1,1} & = & -A_{1,1}^{2}\,,\qquad A_{1,1}^{2}\geq0\,,\\
a_{1,2} & = & A_{1,2}^{2}\geq0\,,\\
a_{2,12} & = & -\frac{1}{2}\,a_{1,3}+\xi^{2}\,,\qquad\xi^{2}\geq0\,,\\
a_{2,22} & = & -3a_{2,13}-\frac{3}{2}\,a_{3,112}\,,\\
a_{2,23} & = & -\frac{3}{4}\,a_{3,122}-\frac{9}{8}\,(g_{3}+g_{4})\,,\\
a_{2,33} & = & -\frac{3}{2}\,a_{3,222}-\frac{9}{2}\,g_{1}-3g_{2}-\frac{9}{2}\,g_{5}-2g_{6}\,,\\
b_{1,1} & = & -B_{1,1}^{2}\,,\qquad B_{1,1}^{2}\geq0\,,\\
b_{1,2} & = & B_{1,2}^{2}\geq0\,,\\
b_{2,12} & = & -\frac{1}{2}\,b_{1,3}+\zeta_{1}^{2}\,,\qquad\zeta_{1}^{2}\geq0\,,\\
b_{2,23} & = & -\frac{3}{4}\,b_{3,122}-\frac{9}{8}\,(h_{3}+h_{4})+\zeta_{3}^{2}\,,\qquad\zeta_{3}^{2}\geq0\,,\\
b_{2,13} & = & -\frac{1}{3}\,b_{2,22}-\frac{1}{2}\,b_{3,112}+\zeta_{2}^{2}\,,\qquad\zeta_{2}^{2}\geq0\,.
\end{eqnarray}
It then turns out that these are sufficient conditions for making also $Z_{1}$ never vanish.

\subsection{Positive $\mu^{2}$ for $^{\forall}X(t)$}

In this process we have assumed that $\mu^{2}$ is positive. However, 
we have to make sure it is. In
fact, we find that on using the previous constraints, the squared
mass for the tensor modes can be rewritten as 
\begin{equation}
\mu^{2}=-\frac{1}{2}m^{2}B_{1,1}^{2}rX^{3}-m^{2}(rB_{1,2}^{2}+A_{1,2}^{2})X^{2}-\frac{1}{2}m^{2}A_{1,1}^{2}X+6m^{2}r\zeta_{2}^{2}+\frac{12}{X}\,m^{2}r\zeta_{3}^{2}+\frac{54}{5X^{2}}\,m^{2}r\zeta_{4}^{2}\,,
\end{equation}
where 
\begin{equation}
b_{2,33}=\zeta_{4}^{2}-\frac{3}{2}\,b_{3,222}-\frac{9}{2}(h_{1}+h_{5})-3h_{2}-2h_{6}\,,
\end{equation}
so that we also need to impose 
\begin{eqnarray}
B_{1,1} & = & 0\,,\\
B_{1,2} & = & 0\,,\\
A_{1,2} & = & 0\,,\\
A_{1,1} & = & 0\,,\\
\zeta_{4}^{2} & \geq & 0\,,
\end{eqnarray}
or 
\begin{equation}
\mu^{2}=6m^{2}r\left(\zeta_{2}^{2}+\frac{2}{X}\,\zeta_{3}^{2}+\frac{9}{5X^{2}}\,\zeta_{4}^{2}\right).
\end{equation}

\subsection{Finiteness of ISW effect}

In the following, we show that the phenomenological criteria so far
are sufficient to guarantee the finiteness of the ISW effect. For
this purpose we use the equations of motion for scalar perturbations
derived in subsection~\ref{subsec:eom-scalarperturbation}.

Since we are interested in the behaviour of dust at late times, we
will consider only one single pressure-less fluid (modeling baryon
and dark matter components). This leads to having effectively only
one kind of matter component, for which the equations of motion reduce
to 
\begin{eqnarray}
E_{\theta} & = & \dot{\delta}_{m}+\theta_{m}-3\dot{\phi}=0\,,\\
E_{\delta\rho/\rho} & = & \dot{\theta}_{m}+aH\,\theta_{m}-k^{2}\psi=0\,.
\end{eqnarray}
Also we consider the subset of the extended theories which satisfy
the conditions $\Xi_{1}=0=\Xi_{2}$, and in particular the model and
the constraints we have found in the previous section leading some
coefficient to vanish, e.g.\ $S_{12}=0$, as to have 
\begin{equation}
E_{E}=S_{1}\dot{\phi}+S_{2}\phi+S_{3}\psi+S_{10}\theta_{m}=0\,.\label{eq:dyn1_aux-1}
\end{equation}
Instead, Eq.\ (\ref{eq:E_and_zeta}) reduces to 
\begin{equation}
E_{E\zeta}=T_{1}\delta\lambda_{V}+T_{2}\phi+T_{3}\psi+T_{4}\delta_{m}+T_{6}\theta_{m}=0\,,
\end{equation}
which, as done before, can be used then to define $\delta\lambda_{V}$
in terms of the other variables. Finally the equation $E_{\delta\lambda}$
simplifies to 
\begin{equation}
E_{\delta\lambda}=U_{1}\phi+U_{2}\delta_{m}+U_{4}\theta_{m}=0\,.\label{eq:dyn_0}
\end{equation}
Now, on taking the time derivative of Eq.\ (\ref{eq:dyn_0}), we
have 
\[
\dot{E}_{\delta\lambda}=(U_{1}+3U_{2})\dot{\phi}+k^{2}U_{4}\psi+\dot{U}_{1}\phi+\dot{U}_{2}\delta_{m}+(\dot{U}_{4}-U_{2}-aHU_{4})\,\theta_{m}=0\,,
\]
where we have replaced the time derivative of the matter fields by
using their own equations of motion. Then we can build up the following
combination of equations of motion 
\begin{eqnarray}
E_{S} & \equiv & (U_{1}+3U_{2})E_{E}-S_{1}\dot{E}_{\delta\lambda}=[(U_{1}+3U_{2})S_{3}-k^{2}U_{4}S_{1}]\psi+[(U_{1}+3U_{2})S_{2}-S_{1}\dot{U}_{1}]\phi\nonumber \\
 & + & [(U_{1}+3U_{2})S_{10}-(\dot{U}_{4}-U_{2}-aHU_{4})S_{1}]\theta_{m}-S_{1}\dot{U}_{2}\delta_{m}=0\,.
\end{eqnarray}
From this last equation, $E_{S}=0$, we find 
\begin{equation}
\psi=F_{\psi}(\phi,\theta_{m},\delta_{m})\,.
\end{equation}
On substituting this expression into $E_{E}=0$, we can solve this equation
for $\dot{\phi}$ as 
\begin{equation}
\dot{\phi}=F_{\dot{\phi}}(\phi,\theta_{m},\delta_{m})\,.
\end{equation}
Then on replacing $\dot{\phi}$ in $E_{\theta}=0$, we can solve it
in terms of $\theta_{m}$, finding 
\begin{equation}
\theta_{m}=F_{\theta}(\phi,\delta_{m},\dot{\delta}_{m})\,,
\end{equation}
from which we also obtain 
\begin{eqnarray}
\psi & = & G_{\psi}(\phi,\delta_{m},\dot{\delta}_{m})\,,\\
\dot{\phi} & = & G_{\dot{\phi}}(\phi,\delta_{m},\dot{\delta}_{m})\,,\\
\dot{\theta}_{m} & = & F_{\dot{\theta}}(\phi,\delta_{m},\dot{\delta}_{m},\ddot{\delta}_{m})\,.
\end{eqnarray}
Then on substituting these expressions in $E_{\delta\rho/\rho}=0$,
we can solve it for $\phi$ as in 
\begin{equation}
\phi=F_{\phi}(\delta_{m},\dot{\delta}_{m},\ddot{\delta}_{m})\,,
\end{equation}
which in turns can be used to set 
\begin{eqnarray}
\psi & = & I_{\psi}(\delta_{m},\dot{\delta}_{m},\ddot{\delta}_{m})\,,\\
\theta_{m} & = & H_{\phi}(\delta_{m},\dot{\delta}_{m},\ddot{\delta}_{m})\,.
\end{eqnarray}

Finally we can substitute these last expressions for $\phi$ and $\theta_{m}$
into Eq.\ (\ref{eq:dyn_0}), $E_{\delta\lambda}=0$, in order to
find a closed differential equation for $\delta_{m}$ of the kind
\begin{equation}
\ddot{\delta}_{m}+A\,\dot{\delta}_{m}+B\,\delta_{m}=0\,.
\end{equation}
Once more, the reason why we can close the dynamical equation of motion
for $\delta_{m}$ is that the theory does not add any new propagating
degree of freedom in the scalar sector. In the high-$k$ regime the
previous equation reduces to 
\begin{equation}
\ddot{\delta}_{m}+aH\,\dot{\delta}_{m}-\frac{3}{2}\,\frac{G_{{\rm eff}}}{G_{N}}\,\Omega_{m}\,a^{2}H^{2}\,\delta_{m}=0\,,
\end{equation}
where $\Omega_{m}=\frac{\rho_{m}}{3\Mpl^{2}H^{2}}$ and the concrete
expression for $G_{{\rm eff}}/G_{N}$ is shown in Appendix \ref{app:Geff_GN}.
This differential equation for $\delta_{m}$ can be used to replace
$\ddot{\delta}_{m}$ in terms of $\delta_{m},\dot{\delta}_{m}$, so
that any scalar perturbation field becomes a function of $\delta_{m},\dot{\delta}_{m}$
only. The result in this section for $G_{{\rm eff}}/G_{N}$, following
a different method, agrees with the one of the previous section, as
expected\footnote{In particular, this result shows that $\delta\rho/\rho=\delta_{m}$
in the high-$k$ regime. It can be proven that this same result holds
also for another gauge invariant combination, the comoving matter
energy density defined as $\delta_{v}=\delta\rho/\rho+3H\,v$, namely
$\delta\rho/\rho=\delta_{v}$.}.

At this point we have found that all the fields (except for $\delta_{m}$
itself) can be written as linear combinations of $\delta_{m}$ and
$\dot{\delta}_{m}$. In particular, we can find the following combination,
whose time derivative affects the ISW effect, namely 
\begin{equation}
\psi_{{\rm ISW}}\equiv\phi+\psi\,.
\end{equation}
We find that in the high-$k$ regime we have 
\begin{equation}
\psi_{{\rm ISW}}=-\frac{3H_{0}^{2}\Omega_{m0}}{k^{2}}\,\Sigma\,\frac{\delta}{a}\,,\label{eq:def_Sigma_ISW}
\end{equation}
where $\Sigma=\Sigma(t)$, $\lim_{m/H\to0}\Sigma=1$, and its denominator
never vanishes for any dynamics of $X(t)$. The general expression
for this model is written in Appendix \ref{app:Geff_GN}.

\subsection{General subclass}

\label{subsec:generalsubclass}

Finally, on putting together all the phenomenological criteria, we
find that the model can be rewritten as 
\begin{eqnarray}
F_{1} & = & c_{4}+\left(\frac{2}{9}[\mathfrak{K}]^{3}-[\mathfrak{K}][\mathfrak{K}^{2}]+[\mathfrak{K}^{3}]\right)a_{1,3}+\left(2[\mathfrak{K}][\mathfrak{K}^{2}]-\frac{10}{9}[\mathfrak{K}]^{3}\right)\xi^{2}+\left(\frac{1}{9}[\mathfrak{K}]^{4}+2[\mathfrak{K}][\mathfrak{K}^{3}]-3[\mathfrak{K}^{2}]^{2}\right)a_{2,13}\nonumber \\
 & + & \left([\mathfrak{K}^{2}]^{2}[\mathfrak{K}]-\frac{1}{18}[\mathfrak{K}]^{5}-\frac{3}{2}[\mathfrak{K}^{2}][\mathfrak{K}^{3}]\right)a_{3,122}+\left([\mathfrak{K}^{2}]^{3}-\frac{[\mathfrak{K}]^{6}}{54}-\frac{3[\mathfrak{K}^{3}]^{2}}{2}\right)a_{3,222}+\left([\mathfrak{K}^{2}]\,[\mathfrak{K}]^{4}-\frac{5}{18}[\mathfrak{K}]^{6}-\frac{9}{2}[\mathfrak{K}^{3}]^{2}\right)g_{1}\nonumber \\
 & + & \left([\mathfrak{K}^{2}]^{2}[\mathfrak{K}]^{2}-\frac{2}{27}[\mathfrak{K}]^{6}-3[\mathfrak{K}^{3}]^{2}\right)g_{2}+\left(\frac{\left(4[\mathfrak{K}]^{3}-9[\mathfrak{K}^{3}]\right)[\mathfrak{K}^{2}]}{4}-\frac{[\mathfrak{K}]^{5}}{4}\right)g_{3}+\left(\frac{\left(36[\mathfrak{K}]^{2}-81[\mathfrak{K}^{2}]\right)[\mathfrak{K}^{3}]}{36}-\frac{[\mathfrak{K}]^{5}}{36}\right)g_{4}\nonumber \\
 & - & \frac{\left([\mathfrak{K}]^{3}-9[\mathfrak{K}^{3}]\right)^{2}g_{5}}{18}+\left([\mathfrak{K}^{2}][\mathfrak{K}^{3}][\mathfrak{K}]-\frac{1}{81}[\mathfrak{K}]^{6}-2[\mathfrak{K}^{3}]^{2}\right)g_{6}-\frac{\left([\mathfrak{K}]^{2}-3[\mathfrak{K}^{2}]\right)^{2}a_{3,112}}{6}\,.\label{eqn:F1-subclass}\\
F_{2} & = & \left(\frac{2}{9}[\mathcal{K}]^{3}-[\mathcal{K}][\mathcal{K}^{2}]+[\mathcal{K}^{3}]\right)b_{1,3}+\left(-\frac{1}{27}[\mathcal{K}]^{4}-\frac{2}{3}[\mathcal{K}][\mathcal{K}^{3}]+[\mathcal{K}^{2}]^{2}\right)b_{2,22}+\left(-\frac{10}{9}[\mathcal{K}]^{3}+2[\mathcal{K}][\mathcal{K}^{2}]\right)\zeta_{1}^{2}\nonumber \\
 & + & \left(2[\mathcal{K}][\mathcal{K}^{3}]-\frac{4}{9}[\mathcal{K}]^{4}\right)\zeta_{2}^{2}+\left(2[\mathcal{K}^{2}][\mathcal{K}^{3}]-\frac{2[\mathcal{K}]^{5}}{15}\right)\zeta_{3}^{2}+\left([\mathcal{K}^{3}]^{2}-\frac{[\mathcal{K}]^{6}}{45}\right)\zeta_{4}^{2}\nonumber \\
 & + & \left([\mathcal{K}^{2}]^{2}[\mathcal{K}]^{2}-\frac{2}{27}[\mathcal{K}]^{6}-3[\mathcal{K}^{3}]^{2}\right)\!h_{2}+\left(\frac{\left(4[\mathcal{K}]^{3}-9[\mathcal{K}^{3}]\right)[\mathcal{K}^{2}]}{4}-\frac{[\mathcal{K}]^{5}}{4}\right)\!h_{3}+\left(\frac{\left(36[\mathcal{K}]^{2}-81[\mathcal{K}^{2}]\right)[\mathcal{K}^{3}]}{36}-\frac{[\mathcal{K}]^{5}}{36}\right)\!h_{4}\nonumber \\
 & + & \left([\mathcal{K}^{2}]^{2}[\mathcal{K}]-\frac{1}{18}[\mathcal{K}]^{5}-\frac{3}{2}[\mathcal{K}^{2}][\mathcal{K}^{3}]\right)b_{3,122}+\left([\mathcal{K}^{2}]^{3}-\frac{[\mathcal{K}]^{6}}{54}-\frac{3[\mathcal{K}^{3}]^{2}}{2}\right)b_{3,222}+\left([\mathcal{K}^{2}]\,[\mathcal{K}]^{4}-\frac{5}{18}[\mathcal{K}]^{6}-\frac{9}{2}[\mathcal{K}^{3}]^{2}\right)h_{1}\nonumber \\
 & + & \left([\mathcal{K}^{2}]\,[\mathcal{K}]^{2}-\frac{2}{9}[\mathcal{K}]^{4}-[\mathcal{K}][\mathcal{K}^{3}]\right)b_{3,112}-\frac{\left([\mathcal{K}]^{3}-9[\mathcal{K}^{3}]\right)^{2}h_{5}}{18}+\left(-\frac{1}{81}[\mathcal{K}]^{6}+[\mathcal{K}^{2}][\mathcal{K}^{3}][\mathcal{K}]-2[\mathcal{K}^{3}]^{2}\right)h_{6}\,.\label{eqn:F2-subclass}
\end{eqnarray}

For this class of models we can see that 
\begin{equation}
\frac{\dot{X}}{NH}=\frac{5\zeta_{1}^{2}X^{3}+10\zeta_{2}^{2}X^{2}+10\zeta_{3}^{2}X+6\zeta_{4}^{2}}{5X^{4}\xi^{2}}\,r-X\,,
\end{equation}
whose dynamics is always well defined. Note two things: 1) a $\Lambda$CDM
profile, i.e.\ $X=X_{0}={\rm constant}$, can always be given for
the background if necessary, and 2) on giving $X(t)$, we find $r(t)$,
or vice versa, on giving $r(t)$, one needs to solve an ODE in order
to find $X(t)$. It is interesting to notice that the Friedmann equation
can then be written as 
\begin{eqnarray}
3\Mpl^{2}H^{2} & = & \sum_{I}\rho_{I}+m^{2}\Mpl^{2}\,(c_{4}-6\xi^{2}X^{3})\,,\\
2\Mpl^{2}\,\frac{\dot{H}}{N} & = & -\sum_{I}(\rho_{I}+P_{I})-6\Mpl^{2}m^{2}\xi^{2}\,\frac{\dot{X}}{NH}\,X^{2}\,,
\end{eqnarray}
which simplify considerably. In this case the equation of state parameter
for the eMTMG component becomes 
\begin{eqnarray}
w_{g} & \equiv & \frac{P_{g}}{\rho_{g}}=\frac{30\left(X^{3}\zeta_{1}^{2}+2\zeta_{2}^{2}X^{2}+2\zeta_{3}^{2}X+\frac{6}{5}\zeta_{4}^{2}\right)r-5X^{2}c_{4}}{5X^{2}\left(c_{4}-6\xi^{2}X^{3}\right)}\,,\\
\rho_{g} & \equiv & m^{2}\Mpl^{2}\,(c_{4}-6\xi^{2}X^{3})\,.
\end{eqnarray}
Also, as already mentioned, the general expression for $G_{{\rm eff}}/G_{N}$
and $\Sigma$ for this general subclass can be found in appendix \ref{app:Geff_GN}.

\subsection{Simple subclass}

\label{subsec:simplesubclass}

We have obtained the general subclass of models in subsection~\ref{subsec:generalsubclass}.
On the other hand, since the observables, $\rho_{g}$, $G_{{\rm eff}}/G_{N}$,
$\Sigma$ and $\mu_{{\rm GW}}^{2}$ depend only on $(c_{4},\xi,\zeta_{1},\zeta_{2},\zeta_{3},\zeta_{4})$
among parameters in (\ref{eqn:F1-subclass})-(\ref{eqn:F2-subclass}),
we can pick up a simple subclass as follows\footnote{As we will see in Appendix \ref{app:selfacce}, the existence of the
self-acceleraring branch requires that $F_{1,[\mathfrak{K}]}+2XF_{1,[\mathfrak{K}^{2}]}+3X^{2}F_{1,[\mathfrak{K}^{3}]}=0$,
which for this subclass leads to imposing $12\xi^{2}X^{2}=0$, a solution
which requires $\xi=0$. On the other hand, we also need to impose
$X^{2}F_{2,[\mathcal{K}]}+2XF_{2,[\mathcal{K}^{2}]}+3F_{2,[\mathcal{K}^{3}]}=0$,
which is solved by setting the condition $5\zeta_{1}^{2}X^{3}+10\zeta_{2}^{2}X^{2}+10\zeta_{3}^{2}X+6\zeta_{4}^{2}=0$.
In this case then $G_{{\rm eff}}/G_{N}=1$, and the phenomenology
reduces to the one of $\Lambda$CDM, except, in general, for a non-zero
value of the graviton mass.} 
\begin{eqnarray}
F_{1} & = & c_{4}+\left(2[\mathfrak{K}][\mathfrak{K}^{2}]-\frac{10}{9}[\mathfrak{K}]^{3}\right)\xi^{2}\,,\label{eq:F1}\\
F_{2} & = & \left(2[\mathcal{K}][\mathcal{K}^{2}]-\frac{10}{9}[\mathcal{K}]^{3}\right)\zeta_{1}^{2}+\left(2[\mathcal{K}][\mathcal{K}^{3}]-\frac{4}{9}[\mathcal{K}]^{4}\right)\zeta_{2}^{2}+\left(2[\mathcal{K}^{2}][\mathcal{K}^{3}]-\frac{2[\mathcal{K}]^{5}}{15}\right)\zeta_{3}^{2}+\left([\mathcal{K}^{3}]^{2}-\frac{[\mathcal{K}]^{6}}{45}\right)\zeta_{4}^{2}\,,\label{eq:F2}
\end{eqnarray}
which has six free parameters instead of 4 for MTMG (or dRGT model).
This subset, at least on FLRW, is sufficiently general in the sense
that it catches the behavior of a more general subclass of models
presented in subsection~\ref{subsec:generalsubclass}.

In principle, one needs to be fitting all the free parameters of the
model against the data, giving then predictions on the graviton mass.
We will study the possibly interesting phenomenology for this theory
in another separate paper.

\section{Conclusion}

\label{sec:concl}

Nowadays, cosmology has reached an astonishingly high level of understanding
of our universe due to more and more precise observations whose number
also grows more and more. However, these data seem to give us a puzzling
scenario regarding the dark sector of our universe. This is not only
due to the long standing problem of understanding the tiny value of the
cosmological constant, for which a complete theoretical explanation
is still unavailable. In fact, recent cosmological observations,
as they reached a percent level of precision (at least for some of
the experiments), show tensions and/or anomalies in the estimation
of cosmological parameters such as the $z=0$ Hubble expansion rate,
$H_{0}$, or the amplitude of matter fluctuations $S_{8}$ in the
context of \foreignlanguage{american}{GR-}$\Lambda$CDM. On the other
hand, the fact that GR has shown to be fully compatible up to now
with local-gravity/astrophysical observations (including the experiments
concerning gravitational waves), seems to leave little space to some
deviations from it. Therefore the puzzle which leads to doubts on
the experiments, or doubts on the analysis/interpretation of the data,
or doubts on the theoretical model used to fit the same data. In particular,
this third possibility opens up a room for new (gravitational) physics
and motivates, for instance, the study of various modified theories
of gravity to address these same tensions.

We have tried, as further explained later, to modify gravity in this
paper under the assumption that the graviton has a nonzero mass. Indeed,
giving a mass to the graviton is a well motivated scenario to consider.
This issue has been attracting the attention of several physicists,
since the first attempt by Fierz and Pauli, back in 1939 \citep{Fierz_Pauli}.
The full nonlinear realization of massive gravity was accomplished
only very recently, which is now known as dRGT theory \citep{dRGT_1}.
Although the theory of dRGT is a valid theory for massive gravity,
nonetheless it was proven that the cosmology of this theory was plagued
with instabilities \citep{dRGT_no_FLRW}, as at least one (out of
the five degrees of freedom) is a ghost (whose mass is in general
below the cutoff of the theory). In~\citep{MTMG:origpap} a new theory
of massive gravity which was constructed as not to have the unstable
mode of dRGT was introduced. This theory, called ``minimal theory
of massive gravity'' (MTMG), is said to be minimal in the sense that,
it does not propagate any degrees of freedom other than the gravitational
waves, which, on the other hand, are massive. This theory shows interesting
phenomenology as it can lower the value of $f\sigma_{8}$~\citep{DeFelice:2016ufg},
since pressure-less fluids can feel weaker gravity, as the effective
gravitational constant is lower than the Newton constant, i.e.\ 
$G_{\text{eff}}<G_{N}$.

However, this modification of $G_{\text{eff}}$ in MTMG 
consists of being a function of time with a pole at $\mu^{2}/H_{\infty}^{2}=2$,
with $\mu^{2}$ being the squared mass of the tensor modes in the theory
and $H_{\infty}$ being the value of the Hubble expansion rate at which
$|G_{{\rm eff}}|\to\infty$, where the background is nonetheless well
defined and equal to $\Lambda$CDM. Then it is clear that the theory
breaks down (its description as a low energy effective theory), for
$H_{\infty}>H_{0}$, i.e.\ $\mu^{2}\geq2H_{0}^{2}$, see e.g.\ \citep{DeFelice:2016ufg,Hagala:2020eax}.
However, if $\mu^{2}<0$ the pole is never encountered, and $G_{{\rm eff}}$
remains a smooth function at all times. The price to pay for this
(in MTMG) is that the gravitational waves are tachyon fields,
possessing a negative, but tiny squared mass. This would lead to a
tachyonic instability for them which is only effective for graviton-kinetic-energy
of order $H_{0}^{2}$ (not visible at astrophysical scales) and a
time of instability of order $H_{0}^{-1}$. Therefore in MTMG, either
we live with tachyonic gravitational waves, or we have to avoid real 
(and larger than $H_{0}$) values for $\mu$. This phenomena do limit
the phenomenological possibilities of the normal branch of the original MTMG.

In this work, we extend the MTMG theory, motivated by the previous
phenomenological behavior, as to remove the negative-squared mass
behavior and, at the same time, any poles in $G_{{\rm eff}}$. We
impose these properties to be valid at any time and for any background
dynamics. By doing this also other observables, such as the ISW field,
will have a smooth evolution. In order to define the extended Minimal Theory of 
Massive Gravity (eMTMG), we first realize that the original MTMG was built as to have the same
cosmological background as dRGT. Then, we allow the new class of theories to
have a general graviton mass term and not only the MTMG/dRGT-like
one. Afterwards, in order to have a theory with only tensor degrees
of freedom in the gravity sector, we implement new constraints as
to remove the unstable modes (already present in dRGT). Now the eMTMG 
leads to a mass term which consists of two functions: a function
$F_{1}$ of $[\mathfrak{K}]$, $[\mathfrak{K}^{2}]$, $[\mathfrak{K}^{3}]$;
and another function $F_{2}$ of $[\mathcal{K}]$, $[\mathcal{K}^{2}]$,
$[\mathcal{K}^{3}]$ (where $[\mathfrak{K}],\dots[\mathcal{K}],\dots$
depend on the three dimensional metric $\gamma_{ij}$ and a fiducial
metric $\tilde{\gamma}_{ij}$). This choice naturally vastly expands
the phenomenology of MTMG in general. After investigating the background
equations of motion, we have studied the tensor mode perturbations,
and found that the two polarizations of the gravitational waves acquire
a nontrivial mass as expected.

Later on, we impose the conditions mentioned above for $\mu^{2}\geq0$
and finiteness of $G_{{\rm eff}}$. We have found that this model
can lead to three possible different phenomenologies, which depend
on two functions $\Xi_{1}$ and $\Xi_{2}$ which, in turn, depend
on $F_{1,2}$ and their derivatives. In fact, we find that if $\Xi_{1}\neq0$,
then $G_{{\rm eff}}=G_{N}$ (evidently MTMG does not belong to this
class). If instead $\Xi_{1}=0$ (or much smaller than $k^{2}/(a^{2}H^{2})$)
but $\Xi_{2}\neq0$, in general the speed of propagation for each
matter field will be modified. Finally for the subclass of theories
for which $\Xi_{1}=0=\Xi_{2}$, matter component has the standard speed
of propagation, whereas dust acquires a nontrivial $G_{{\rm eff}}/G_{N}$.
MTMG belongs to this last class. For this last class we proceed to
impose the conditions $\mu^{2}\geq0$ and $G_{{\rm eff}}<\infty$,
and we give an explicit example which satisfies these constraints
at all times for any background dynamics.

In this last case, the expression for $G_{\text{eff}}/G_{N}$
is explicitly given in Appendix~\ref{app:Geff_GN}. It is interesting
to notice that the mass squared of the graviton could be vanishing,
whereas $G_{{\rm eff}}\neq G_{N}$. This is due to the fact that at
linear level, the contributions to $G_{{\rm eff}}$ come from the
would-be-unstable propagating scalar mode of dRGT which in this minimal
theory is non-dynamical and as such can be integrated out, leading
though to nonstandard modifications to the coefficients of the linear
perturbation equations of motion.

We have extended the study of finiteness to other linear perturbation observables
as to see how their late time dynamics are affected. In particular,
we have looked at the observable which describes the ISW-galaxy correlation
effects. Indeed we find that imposing $\mu^{2}\geq0$ and $G_{{\rm eff}}<\infty$
automatically leads to the absence of poles for such observable.

The result of this work is interesting since it provides a set of eMTMG 
which, like GR, lead to cosmological
observables which are always well defined, no matter which dynamics
the background might have. These requirements can turn to be crucial
in a world which has to deal with a weak gravity description of large-scale
gravitational interactions. We think these minimal models could have
an interesting phenomenology leading to new possibilities for a massive
graviton to play a non-trivial role in our physical world.

\begin{acknowledgments}
The work of A.D.F.\ was supported by Japan Society for the Promotion
of Science Grants-in-Aid for Scientific Research No.\ 20K03969. The
work of S.M.\ was supported by Japan Society for the Promotion of
Science Grants-in-Aid for Scientific Research No.\ 17H02890, No.\ 17H06359,
and by World Premier International Research Center Initiative, MEXT,
Japan. The work of M.C.P.\ was supported by the Japan Society for
the Promotion of Science Grant-in-Aid for Scientific Research No.\ 17H06359. 
\end{acknowledgments}

\appendix

\section{Variational fomulae}

\label{app:variations}

We find it useful to write down some identities which are to be used
when we find the equations of motion of the theory\footnote{In the following we will make use of the identity $\delta\{{\rm Tr}[\sqrt{X}^{n}]\}=\frac{n}{2}\,{\rm Tr}\bigl[\sqrt{X}^{n-2}\delta X\bigr]$.}
\begin{eqnarray}
\delta\mathcal{K}^{i}{}_{i} & = & \frac{1}{2}\,\mathfrak{K}^{i}{}_{j}\tilde{\gamma}^{jl}\delta\gamma_{li}\,,\\
\partial_{t}\mathcal{K}^{i}{}_{i} & = & \frac{1}{2}\,\mathfrak{K}^{i}{}_{j}\dot{\tilde{\gamma}}^{jl}\gamma_{li}=-M\,\mathfrak{K}^{i}{}_{j}\bar{\zeta}^{j}{}_{i}\,,\\
\delta(\mathcal{K}^{i}{}_{j}\mathcal{K}^{j}{}_{i}) & = & \tilde{\gamma}^{ij}\delta\gamma_{ij}\,,\\
\partial_{t}(\mathcal{K}^{i}{}_{j}\mathcal{K}^{j}{}_{i}) & = & \dot{\tilde{\gamma}}^{il}\gamma_{li}=-2M\,\bar{\zeta}^{i}{}_{j}\,\tilde{\gamma}^{jl}\,\gamma_{li}\,,\\
\delta(\mathcal{K}^{i}{}_{j}\mathcal{K}^{j}{}_{k}\mathcal{K}^{k}{}_{i}) & = & \frac{3}{2}\,\mathcal{K}^{i}{}_{j}\tilde{\gamma}^{jl}\delta\gamma_{li}\,,\\
\partial_{t}(\mathcal{K}^{i}{}_{j}\mathcal{K}^{j}{}_{k}\mathcal{K}^{k}{}_{i}) & = & \frac{3}{2}\,\mathcal{K}^{i}{}_{j}\dot{\tilde{\gamma}}^{jl}\gamma_{li}=-3M\,\mathcal{K}^{i}{}_{j}\,\bar{\zeta}^{j}{}_{l}\,\tilde{\gamma}^{lk}\,\gamma_{ki}\,,
\end{eqnarray}
where a $\delta$ represents the variation with respect to a dynamical
field, and $\partial_{t}$ the time-derivative of the explicitly time
dependent fields.

Now let us turn our attention to the analogue properties of the other
squared-root matrix, namely $\mathfrak{K}^{i}{}_{j}$. Then we find
\begin{eqnarray}
\delta\mathfrak{K}^{i}{}_{i} & = & \frac{1}{2}\,\mathcal{K}^{i}{}_{j}\delta\gamma^{jl}\tilde{\gamma}_{li}=-\frac{1}{2}\,\mathcal{K}^{i}{}_{j}\gamma^{kl}\tilde{\gamma}_{li}\gamma^{jm}\delta\gamma_{mk}=-\frac{1}{2}\,\mathfrak{K}^{i}{}_{j}\,\gamma^{jk}\delta\gamma_{ki}\,,\\
\partial_{t}\mathfrak{K}^{i}{}_{i} & = & \frac{1}{2}\,\mathcal{K}^{i}{}_{j}\gamma^{jl}\dot{\tilde{\gamma}}_{il}=M\,\tilde{\gamma}_{ij}\mathcal{K}^{i}{}_{l}\gamma^{kl}\,\bar{\zeta}^{j}{}_{k}\,,\\
\delta(\mathfrak{K}^{i}{}_{j}\mathfrak{K}^{j}{}_{i}) & = & \delta\gamma^{ij}\tilde{\gamma}_{ij}=-\gamma^{ij}\tilde{\gamma}_{jl}\gamma^{lm}\delta\gamma_{mi}\,,\\
\partial_{t}(\mathfrak{K}^{i}{}_{j}\mathfrak{K}^{j}{}_{i}) & = & 2M\gamma^{ij}\tilde{\gamma}_{jl}\,\bar{\zeta}^{l}{}_{i}\,,\\
\delta(\mathfrak{K}^{i}{}_{j}\mathfrak{K}^{j}{}_{k}\mathfrak{K}^{k}{}_{i}) & = & \frac{3}{2}\,\mathfrak{K}^{i}{}_{j}\tilde{\gamma}_{li}\delta\gamma^{jl}=-\frac{3}{2}\,\mathfrak{K}^{i}{}_{j}\tilde{\gamma}_{li}\gamma^{ml}\gamma^{jk}\delta\gamma_{km}\,,\\
\partial_{t}(\mathfrak{K}^{i}{}_{j}\mathfrak{K}^{j}{}_{k}\mathfrak{K}^{k}{}_{i}) & = & \frac{3}{2}\,\mathfrak{K}^{i}{}_{j}\gamma^{jl}\dot{\tilde{\gamma}}_{li}=3M\,\mathfrak{K}^{i}{}_{j}\gamma^{jl}\tilde{\gamma}_{lm}\,\bar{\zeta}^{m}{}_{i}\,.
\end{eqnarray}

\section{MTMG subcase}

\label{app:MTMG-subcase}

In the theory of MTMG, a subclass of the eMTMG, we have for
the precursor part of the Lagrangian the following structure 
\begin{eqnarray}
\mathcal{L}_{{\rm MTMG}} & \ni & \frac{m^{2}\Mpl^{2}}{2}\,N\,[-c_{1}\sqrt{\tilde{\gamma}}-c_{2}\sqrt{\tilde{\gamma}}[\mathcal{K}]-c_{3}\sqrt{\gamma}\,[\mathfrak{K}]-c_{4}\sqrt{\gamma}]\nonumber \\
 & = & -\frac{m^{2}\Mpl^{2}}{2}\,N\,\sqrt{\gamma}\left[c_{1}\frac{\sqrt{\tilde{\gamma}}}{\sqrt{\gamma}}+c_{2}[\mathcal{K}]\frac{\sqrt{\tilde{\gamma}}}{\sqrt{\gamma}}+c_{3}[\mathfrak{K}]+c_{4}\right],
\end{eqnarray}
where, by definition
\begin{eqnarray*}
\mathfrak{K}^{a}{}_{b}\mathfrak{K}^{b}{}_{c} & = & \gamma^{ab}\tilde{\gamma}_{bc}\,,\\{}
[\mathcal{K}] & = & [\mathfrak{K}^{-1}]\,,
\end{eqnarray*}
so that 
\begin{equation}
\det(\mathfrak{K})^{2}=\frac{\tilde{\gamma}}{\gamma}\,,\qquad\frac{\sqrt{\tilde{\gamma}}}{\sqrt{\gamma}}=\det(\mathfrak{K})\,,
\end{equation}
supposing that $\det(\mathfrak{K})>0$. By using the Cayley-Hamilton
(CH) theorem\footnote{For a three-dimensional matrix, $A$, one has: 
\[
A^{3}-{\rm tr}(A)\,A^{2}+\frac{1}{2}\,[({\rm tr}A)^{2}-{\rm tr}(A^{2})]\,A-\det(A)\,I_{3}=0\,,
\]
out of which we can take the trace or multiply it by $A^{-1}$ to
find new useful relations.} we have 
\begin{equation}
[\mathfrak{K}^{3}]-[\mathfrak{K}][\mathfrak{K}^{2}]+\frac{1}{2}\,([\mathfrak{K}]^{2}-[\mathfrak{K}^{2}])\,[\mathfrak{K}]-3\det(\mathfrak{K})=0\,,
\end{equation}
so that 
\begin{equation}
\det(\mathfrak{K})=\frac{1}{3}\,[\mathfrak{K}^{3}]-\frac{1}{2}\,[\mathfrak{K}][\mathfrak{K}^{2}]+\frac{1}{6}\,[\mathfrak{K}]^{3}\,.
\end{equation}
We also have from the CH theorem that: 
\begin{equation}
[\mathfrak{K}^{2}]-[\mathfrak{K}]^{2}+\frac{3}{2}\,([\mathfrak{K}]^{2}-[\mathfrak{K}^{2}])-\det(\mathfrak{K})\,[\mathcal{K}]=0\,,
\end{equation}
and 
\begin{equation}
[\mathcal{K}]\det(\mathfrak{K})=\frac{1}{2}\,([\mathfrak{K}]^{2}-[\mathfrak{K}^{2}])\,.
\end{equation}
Since 
\begin{equation}
\mathcal{L}_{{\rm MTMG}}\ni-\frac{m^{2}\Mpl^{2}}{2}N\sqrt{\gamma}\left[c_{1}\det(\mathfrak{K})+c_{2}\,[\mathcal{K}]\det(\mathfrak{K})+c_{3}[\mathfrak{K}]+c_{4}\right],
\end{equation}
we have that for MTMG: 
\begin{equation}
F_{1}^{{\rm MTMG}}=c_{1}\left(\frac{1}{3}\,[\mathfrak{K}^{3}]-\frac{1}{2}[\mathfrak{K}][\mathfrak{K}^{2}]+\frac{1}{6}\,[\mathfrak{K}]^{3}\right)+\frac{1}{2}\,c_{2}\,([\mathfrak{K}]^{2}-[\mathfrak{K}^{2}])+c_{3}\,[\mathfrak{K}]+c_{4}\,.
\end{equation}
Along the same lines one finds 
\begin{eqnarray}
\mathcal{L}_{{\rm MTMG}} & \ni & \frac{m^{2}\Mpl^{2}}{2}M[-c_{1}[\mathcal{K}]\sqrt{\tilde{\gamma}}-\frac{1}{2}c_{2}\sqrt{\tilde{\gamma}}\left([\mathcal{K}]^{2}-[\mathcal{K}^{2}]\right)-c_{3}\sqrt{\gamma}]\nonumber \\
 & = & -\frac{m^{2}\Mpl^{2}}{2}M\sqrt{\tilde{\gamma}}\left[c_{1}[\mathcal{K}]+\frac{1}{2}\,c_{2}\left([\mathcal{K}]^{2}-[\mathcal{K}^{2}]\right)+c_{3}\,\frac{\sqrt{\gamma}}{\sqrt{\tilde{\gamma}}}\right]\,,
\end{eqnarray}
where 
\begin{equation}
\mathfrak{\mathcal{K}}^{a}{}_{b}\mathcal{K}^{b}{}_{c}=\tilde{\gamma}^{ab}\,\gamma_{bc}\,,
\end{equation}
so that 
\begin{equation}
\det(\mathcal{K})^{2}=\frac{\gamma}{\tilde{\gamma}}\,,
\end{equation}
or 
\begin{equation}
\frac{\sqrt{\gamma}}{\sqrt{\tilde{\gamma}}}=\det(\mathcal{K})\,,
\end{equation}
supposing that $\det(\mathcal{K})>0$. Then on using once more the
CH theorem one finds 
\begin{equation}
\det(\mathcal{K})=\frac{1}{3}\,[\mathcal{K}^{3}]-\frac{1}{2}\,[\mathcal{K}][\mathcal{K}^{2}]+\frac{1}{6}\,[\mathcal{K}]^{3}\,,
\end{equation}
and, finally, that 
\begin{equation}
F_{2}^{{\rm MTMG}}=c_{1}\,[\mathcal{K}]+\frac{1}{2}\,c_{2}\left([\mathcal{K}]^{2}-[\mathcal{K}^{2}]\right)+c_{3}\left(\frac{1}{3}\,[\mathcal{K}^{3}]-\frac{1}{2}\,[\mathcal{K}][\mathcal{K}^{2}]+\frac{1}{6}\,[\mathcal{K}]^{3}\right).
\end{equation}

\section{Self-accelerating branch}

\label{app:selfacce}

Let us once more consider the nontrivial constraint equation Eq.\ (\ref{eq:constr_FLRW}),
that we rewrite here for later convenience

\begin{equation}
H\,\frac{M}{N}\left(X^{2}F_{2,[\mathcal{K}]}+2XF_{2,[\mathcal{K}^{2}]}+3F_{2,[\mathcal{K}^{3}]}\right)=\biggl(\frac{\dot{X}}{N}+HX\biggr)\left(F_{1,[\mathfrak{K}]}+2XF_{1,[\mathfrak{K}^{2}]}+3X^{2}F_{1,[\mathfrak{K}^{3}]}\right).\label{eq:constr_FLRW-1}
\end{equation}
We can define a self-accelerating branch for these extended minimal
models, as the solution of this constraint which does not fix the
ratio $M/N$. For this to happen we require 
\begin{equation}
X^{2}F_{2,[\mathcal{K}]}+2XF_{2,[\mathcal{K}^{2}]}+3F_{2,[\mathcal{K}^{3}]}=0\,,\label{eq:SA1}
\end{equation}
which is an algebraic equation for $X$. In particular, this equation
implies that $X=X_{0}={\rm constant}$. Since, from our assumptions
$X_{0}\neq0$, in general, Eq.\ (\ref{eq:constr_FLRW-1}) also leads
to 
\begin{equation}
F_{1,[\mathfrak{K}]}+2XF_{1,[\mathfrak{K}^{2}]}+3X^{2}F_{1,[\mathfrak{K}^{3}]}=0\,.\label{eq:SA2}
\end{equation}
Viceversa, if we assume Eq.\ (\ref{eq:SA2}) holding true, then since
we assume that $HM/N$ does not vanish, we are left to impose that
also Eq.\ (\ref{eq:SA1}) needs to be satisfied. Then both Eqs.\ (\ref{eq:SA1})
and (\ref{eq:SA2}) must hold at the same time, meaning that $X_{0}$
has to be a solution for both these equations. In this case, we will
name this possibility as the self accelerating solution. This solution
might not exist for all possible $F_{1,2}$ functions, but there will
be subclass of theories admitting its presence. In particular MTMG
is one of them.

For the self accelerating branch, as defined here, we find that both
the background and the scalar/vector linear perturbation equations
behave exactly as in General Relativity, and in particular, $\frac{1}{2}\,\Mpl^{2}\,m^{2}\,F_{1}$
reduce to an effective cosmological constant contribution to the total
matter sector. In summary, for this solution, all the phenomenology
(up to linear perturbations in cosmology) coincide with GR except
for the tensor modes which acquire a nonzero mass (possibly time dependent).

\section{Full expression of $G_{{\rm eff}}/G_{N}$}

\label{app:Geff_GN}

In the following we give a full expression for $G_{{\rm eff}}/G_{N}$
which can be written as 
\begin{eqnarray}
\frac{G_{{\rm eff}}}{G_{N}} & = & \frac{1}{\Delta}\left\{ -6750\xi^{2}X^{3}\left(X^{3}\zeta_{1}^{2}+2X^{2}\zeta_{2}^{2}+2X\zeta_{3}^{2}+\frac{6}{5}\zeta_{4}^{2}\right)\left(X^{5}\xi^{2}-X^{3}r\zeta_{1}^{2}-2X^{2}r\zeta_{2}^{2}-2Xr\zeta_{3}^{2}-\frac{6}{5}r\zeta_{4}^{2}\right)\right.\nonumber \\
 & \times & \left(X^{3}\zeta_{1}^{2}+\frac{8}{3}X^{2}\zeta_{2}^{2}+\frac{10}{3}X\zeta_{3}^{2}+\frac{12}{5}\zeta_{4}^{2}\right)m^{4}+3375H^{2}m^{2}\left[\xi^{2}\zeta_{1}^{4}\left(\Omega_{m}+\frac{2}{3}\right)X^{11}+\frac{14\zeta_{2}^{2}\xi^{2}\left(\Omega_{m}+\frac{6}{7}\right)\zeta_{1}^{2}X^{10}}{3}\right.\nonumber \\
 & + & \frac{16\left[\left(\Omega_{m}+\frac{4}{3}\right)\zeta_{2}^{4}+\zeta_{1}^{2}\zeta_{3}^{2}\left(\Omega_{m}+\frac{5}{6}\right)\right]\xi^{2}X^{9}}{3}+\left(\left(12\left(\Omega_{m}+\frac{14}{9}\right)\zeta_{3}^{2}\zeta_{2}^{2}+\frac{18\zeta_{4}^{2}\left(\Omega_{m}+\frac{2}{3}\right)\zeta_{1}^{2}}{5}\right)\xi^{2}-\frac{8r\zeta_{1}^{4}\zeta_{2}^{2}}{9}\right)X^{8}\nonumber \\
 & + & \left(\left(8\zeta_{4}^{2}\left(\Omega_{m}+\frac{8}{5}\right)\zeta_{2}^{2}+\frac{20\zeta_{3}^{4}\left(\Omega_{m}+2\right)}{3}\right)\xi^{2}-\frac{8r\zeta_{1}^{2}\left(\zeta_{1}^{2}\zeta_{3}^{2}+6\zeta_{2}^{4}\right)}{9}\right)X^{7}\nonumber \\
 & + & \left(\frac{44\left(\Omega_{m}+\frac{74}{33}\right)\zeta_{4}^{2}\xi^{2}\zeta_{3}^{2}}{5}-\frac{128r\zeta_{2}^{2}\left(\zeta_{1}^{2}\zeta_{3}^{2}+\frac{\zeta_{2}^{4}}{2}\right)}{9}\right)X^{6}-\frac{144r\left(\zeta_{1}^{2}\zeta_{4}^{4}+\frac{260}{27}\zeta_{2}^{2}\zeta_{3}^{2}\zeta_{4}^{2}+\frac{250}{81}\zeta_{3}^{6}\right)X^{3}}{25}\nonumber \\
 & + & \left(\frac{72\zeta_{4}^{4}\xi^{2}\left(\Omega_{m}+\frac{8}{3}\right)}{25}-\frac{128\left(\zeta_{1}^{2}\zeta_{2}^{2}\zeta_{4}^{2}+\frac{5}{4}\zeta_{1}^{2}\zeta_{3}^{4}+\frac{10}{3}\zeta_{2}^{4}\zeta_{3}^{2}\right)r}{15}\right)X^{5}-\frac{224r\left(\zeta_{1}^{2}\zeta_{3}^{2}\zeta_{4}^{2}+\frac{9}{7}\zeta_{2}^{4}\zeta_{4}^{2}+\frac{55}{21}\zeta_{2}^{2}\zeta_{3}^{4}\right)X^{4}}{15}\nonumber \\
 & - & \left.\frac{512r\left(\zeta_{2}^{2}\zeta_{4}^{2}+\frac{15\zeta_{3}^{4}}{8}\right)\zeta_{4}^{2}X^{2}}{25}-\frac{704Xr\zeta_{3}^{2}\zeta_{4}^{4}}{25}-\frac{864r\zeta_{4}^{6}}{125}\right]\nonumber \\
 &  & +\left.10H^{4}X^{2}\left(15X^{3}\zeta_{1}^{2}+40X^{2}\zeta_{2}^{2}+50X\zeta_{3}^{2}+36\zeta_{4}^{2}\right)^{2}\right\} ,\\
\Delta & \equiv & 2250X^{2}\left(X^{3}\xi^{2}\left(X^{3}\zeta_{1}^{2}+2X^{2}\zeta_{2}^{2}+2X\zeta_{3}^{2}+\frac{6}{5}\zeta_{4}^{2}\right)m^{2}+\frac{H^{2}\left(15X^{3}\zeta_{1}^{2}+40X^{2}\zeta_{2}^{2}+50X\zeta_{3}^{2}+36\zeta_{4}^{2}\right)}{15}\right)^{2},
\end{eqnarray}
where we can explicitly see that the denominator $\Delta$ never vanishes.
We give in the following the general expression for $\Sigma$, defined
in Eq.\ (\ref{eq:def_Sigma_ISW}), which can be written as 
\begin{equation}
\Sigma=\frac{1}{2}\,\frac{G_{{\rm eff}}}{G_{N}}+\frac{H^{2}\left(15X^{3}\zeta_{1}^{2}+40X^{2}\zeta_{2}^{2}+50X\zeta_{3}^{2}+36\zeta_{4}^{2}\right)}{30X^{3}\xi^{2}\left(X^{3}\zeta_{1}^{2}+2X^{2}\zeta_{2}^{2}+2X\zeta_{3}^{2}+\frac{6}{5}\zeta_{4}^{2}\right)m^{2}+2H^{2}\left(15X^{3}\zeta_{1}^{2}+40X^{2}\zeta_{2}^{2}+50X\zeta_{3}^{2}+36\zeta_{4}^{2}\right)}\,,
\end{equation}
which never blows up to infinity for any dynamics of $X(t)$, and
still reduces to unity when $m/H\to0$, i.e.\ at early times.

\section{Case with massless gravitational waves}

\label{app:masslesscase}

For the special symmetric\footnote{We name it ``symmetric'' as in this case Eqs.\ (\ref{eq:F1}) and
(\ref{eq:F2}) mirror each other (a cosmological constant in the fiducial
sector can always be added without modifying any bit of the theory).} model $\zeta_{2}=\zeta_{3}=\zeta_{4}=0$, the tensor modes become
effectively massless on the FLRW background, for any $X(t)$. It is
interesting to note that even if $\mu^{2}$ vanishes, still we might
have non-trivial dynamics in the scalar sector as 
\begin{equation}
\frac{G_{{\rm eff}}}{G_{N}}=\frac{1-3\xi^{4}Y^{2}X^{6}+3X^{3}\left(XYr\zeta_{1}^{2}+\frac{\Omega_{m}}{2}+\frac{1}{3}\right)Y\,\xi^{2}}{\left(Y\,\xi^{2}X^{3}+1\right)^{2}}\,,\qquad Y\equiv\frac{m^{2}}{H^{2}}\,,\quad\Omega_{m}\equiv\frac{\rho_{m}}{3\Mpl^{2}H^{2}}\,,\quad{\rm for}\;\zeta_{2}=\zeta_{3}=\zeta_{4}=0\,.
\end{equation}
which can be still less than unity (but positive) today.

The background function $\Sigma$ defined in (\ref{eq:def_Sigma_ISW})
and shown in Appendix \ref{app:Geff_GN} reduces to 
\begin{equation}
\Sigma=\frac{2-3\xi^{4}Y^{2}X^{6}+3X^{3}Y\left(XYr\zeta_{1}^{2}+\frac{\Omega_{m}}{2}+\frac{2}{3}\right)\xi^{2}}{2\left(Y\,\xi^{2}X^{3}+1\right)^{2}}\,,\qquad Y\equiv\frac{m^{2}}{H^{2}}\,,\quad\Omega_{m}\equiv\frac{\rho_{m}}{3\Mpl^{2}H^{2}}\,,\quad{\rm for}\;\zeta_{2}=\zeta_{3}=\zeta_{4}=0\,.
\end{equation}

Here, we have mentioned this choice for the parameters only as to
give already a non-trivial example despite vanishing mass for gravitational
waves. In the rest of the present paper we shall mainly consider the
more general cases, i.e. those shown in subsections~\ref{subsec:generalsubclass}
or \ref{subsec:simplesubclass}, with massive gravitational waves.

\bibliographystyle{apsrev4-2}
\bibliography{extMTMG_biblio}

\end{document}